\documentclass[12pt]{aastex63}
\usepackage{rotating}
\usepackage{longtable}
\usepackage{graphicx}
\usepackage{txfonts}
\usepackage{verbatim}
\usepackage{enumitem}

\usepackage{natbib,twoopt}
\usepackage{color}
\usepackage{threeparttablex}

\newcommand{\lsun}{log ($L/L_{\odot})\,$}
\newcommand{\msun}{$M/M_{\odot}\,$}

\defcitealias{Desomma2020}{DS20}
\defcitealias{Desomma2022}{DS22}
\defcitealias{Marconi2015}{M15}
\defcitealias{Marconi2024}{M24}

\received{24 Oct 2024}
\revised{29 Oct 2024}
\accepted{1 Nov 2024}
\submitjournal{ApJ}
\shorttitle{SPECTRUM-I}
\shortauthors{G. De Somma et al. }
\graphicspath{{./}{figures/}}
\begin{document}

\title{Stellar Pulsation and Evolution: a Combined Theoretical Renewal and Updated Models (SPECTRUM) - I: Updating radiative opacities for pulsation models of Classical Cepheid and RR-Lyrae.}
\correspondingauthor{Giulia De Somma}
\email{giulia.desomma@inaf.it , gdesomma@na.infn.it}

\author{Giulia De Somma}
\affiliation{ INAF-Osservatorio astronomico di Capodimonte \\
Via Moiariello 16 \\
80131 Napoli, Italy}
\affiliation{INAF-Osservatorio Astronomico d'Abruzzo \\
Via Maggini sn \\
64100 Teramo, Italy\\}
\affiliation{Istituto Nazionale di Fisica Nucleare (INFN) - Sez. di Napoli\\
Compl. Univ.di Monte S. Angelo, Edificio G, Via Cinthia \\
I-80126, Napoli, Italy}

\author{Marcella Marconi}
\affiliation{ INAF-Osservatorio astronomico di Capodimonte \\
Via Moiariello 16 \\
80131 Napoli, Italy}

\author{Santi Cassisi}
\affiliation{INAF-Osservatorio Astronomico d'Abruzzo \\
Via Maggini sn \\
64100 Teramo, Italy\\}
\affiliation{Istituto Nazionale di Fisica Nucleare (INFN) - Sezione di Pisa\\
Universit\'a di Pisa, Largo Pontecorvo 3\\
56127 Pisa, Italy\\}

\author{Roberto Molinaro}
\affiliation{ INAF-Osservatorio astronomico di Capodimonte \\
Via Moiariello 16 \\
80131 Napoli, Italy}

\begin{abstract}
\noindent
Pulsating stars are universally recognized as precise distance indicators and tracers of stellar populations. Their variability, combined with well-defined relationships between pulsation properties and intrinsic evolutionary parameters such as luminosity, mass, and age, makes them essential for understanding galactic evolution and retrieving star formation histories. Therefore, accurate modeling of pulsating stars is crucial for using them as standard candles and stellar population tracers.
This is the first paper in the "Stellar Pulsation and Evolution: a Combined Theoretical Renewal and Updated Models" (SPECTRUM) project, which aims to present an update of Stellingwerf's hydrodynamical pulsation code, by adopting the latest radiative opacity tables commonly used in stellar evolution community.
We assess the impact of this update on pulsation properties, such as periods, instability strip topology, and light curve shapes, as well as on Period Wesenheit and Period-Luminosity relations for Classical Cepheids and RR Lyrae stars, comparing the results with those derived using older opacity data.
Our results indicate that the opacity update introduces only minor changes: instability strip boundary locations shift by no more than $100K$ in effective temperature, and pulsation periods vary within $1\sigma$ compared to previous evaluations. Light curves exhibit slight differences in shape and amplitude. Consequently, the theoretical calibration of the Cepheid or RRL-based extragalactic distance scale remains largely unaffected by the opacity changes.
However, achieving consistency in opacity tables between stellar evolution and pulsation codes is a significant step toward a homogeneous and self-consistent stellar evolution and pulsation framework.

%250 parole 

\end{abstract}

%% Keywords should appear after the \end{abstract} command. 
%% See the online documentation for the full list of available subject
%% keywords and the rules for their use.

\section{Introduction} \label{sec:intro}
Pulsating stars exhibit periodic variations in their luminosity and radius. The relation connecting their mean density to the pulsation period enables the link between an observed pulsation property and the intrinsic stellar parameters, reflecting the evolutionary stage of the investigated pulsator. An accurate formulation of this relationship is currently derived through the development of hydrodynamical models of the pulsation phenomenon. The connection between pulsation and evolutionary properties is the basis for using pulsating stars as stellar population tracers. Furthermore, the possibility to relate the pulsation period to intrinsic luminosity and color makes these objects excellent standard candles. This holds for several classes of pulsating stars including Classical, Type II, and Anomalous Cepheids, as well as RR Lyrae stars.

The reliability of stellar pulsation models strongly relies on the accuracy of the physical ingredients adopted in their computation, such as radiative opacity and thermodynamic properties, notably the equation of state.  

In particular, the stellar opacity plays a key role in driving pulsation mechanisms.
Dating back to S.A.Zhevakin (1963) it is well known that the stellar oscillation phenomenon is essentially a 'Valve effect' \citep[][]{Zhevakin1963}, with the $\kappa$ and $\gamma$ mechanisms, related to the variation of the opacity and the adiabatic exponents, respectively, in the regions where ionization of the most abundant elements of a stellar envelope (H, He and He+) occurs. Indeed the opacity is higher (and the adiabatic exponents are smaller) in partially ionized regions than in completely ionized ones, thus producing energy trapping during the contraction phase, released as excess energy during the subsequent expansion phase which is converted into pulsation work.

According to these mechanisms, the origin of the instability lies in the external layers of the star, where partial ionization occurs, and the 'Valve effect' only activates in a specific temperature range. 

Historical evidence of the importance of updating opacity tables has been highlighted in the context of the well-known Cepheid mass discrepancy problem. This issue arises because the masses calculated using the three main techniques, evolutionary models, pulsation models, and binary systems, do not always agree. Specifically, the masses derived from pulsation models used to be consistently lower (by about 20-40\%) than those derived from evolutionary models \citep[see e.g.][]{Cogan1970, Rodgers1970, Stobie1969}. This discrepancy has posed a significant challenge for both evolution and pulsation theory. In the early '90s, the availability of updated Livermore OPAL radiative opacities \citep[][]{Rogers_Iglesias1992} helped to significantly reduce the mass discrepancy to 10-15\% \citep[see e.g.][]{Moskalik1992, Keller2002}. However, as further improvements in stellar opacity evaluations do not seem to offer a viable solution to the residual discrepancy, additional physical processes, such as the inclusion of mass loss and core convective overshooting, are required to fully resolve the mismatch between pulsation and evolutionary Classical Cepheid (CC) masses \citep[][]{Bono2006, Cassisi_Salaris_2011, Keller2008, Marconi2013, Marconi2017}. Nonetheless, it remains evident that accurate and reliable prescriptions about the opacity behavior of stellar matter are crucial ingredients for modeling both stellar pulsation and evolutionary properties.

On the other hand, over the last decade, significant advancements have been made in updating and refining predictions about both the thermodynamic and opacitive properties of stellar matter in thermal regimes suitable for low- and intermediate-mass pulsating stars. However, these improvements have not yet been fully incorporated into pulsation models. Therefore, it is crucial to test how this updated physical framework impacts the properties of various classes of pulsating stars.

At the same time, most work on modeling pulsating stars has relied on the building of grids of models whose initial parameters - in terms of luminosity and effective temperature - have been obtained from evolutionary stellar models, but without taking into account how they change along an evolutionary track as a consequence of the change in the structural properties along the evolution of the star. The work presented by \citet[][]{Paxton2019} is the only available exception to this approach which was traditionally due to the extremely demanding computational time required by the facilities available in the past. Nowadays, the availability of high-performance computers and more efficient computing algorithms has led to a paradigm shift, enabling the simultaneous modeling of both stellar evolutionary and pulsational properties. This can be achieved by computing the pulsational properties of a stellar model {\sl on the fly} as it evolves through the pulsational instability strip (IS).

Given the aforementioned achievements and the significant impact expected from several ongoing observational surveys, such as Gaia, APOGEE, and TESS \citep[][]{Brown2018, Majewski2017, Perez2016, Ricker2015, Stassun2018}, it is evident that a revolutionized and more accurate theoretical scenario is necessary. This need led us to develop a new project, named {\sl SPECTRUM} (The Stellar Pulsation and Evolution: a Combined Theoretical Renewal and Updated Models). 

One of the main aims of the SPECTRUM project is to update the physics in our pulsation code. While changes to the equation of state are not expected to produce significant variations, as a preliminary study by \citet{Petroni2003} suggested, updating stellar opacity is a crucial step. The latest opacity release needs to be evaluated through the computation of extensive sets of pulsating models for both CC and RRL stars. The goal is to assess how the updated physical framework could affect the predicted properties of these stars and to develop an updated, self-consistent theoretical framework suitable for analyzing and interpreting data from observational surveys such as Gaia DR4, JWST, and the Vera Rubin LSST survey \citep[][]{Abell2009, Brown2016, Brown2018, Brown2021, Gardner2006, Ivezic2019, Prusti2016, Rigby2023} .

At the same time, this project will explore the possibility of simultaneously computing the evolutionary and pulsational properties of variable stars, by integrating pulsation computations into an updated evolutionary code. This integration will enable the self-consistent modeling of variable stars without separate computations for evolutionary and pulsational properties.

This paper is devoted to presenting the project and discussing the effect of implementing more updated opacity tables into the adopted pulsation code on model predictions of the pulsation properties of both CCs and RRLs, which are taken as representative examples of intermediate- and low-mass pulsating stars.

In the present work, we have built a new dataset of 1D pulsation models for various types of variable stars. However, another potential avenue for advancing pulsation hydrodynamic models is through multi-dimensional (2D-3D) calculations \citep[][]{Geroux2011}. These models hold promise for future research, given the theoretical challenge of accurately modeling convection—a fundamentally 3D phenomenon—in both pulsation and evolutionary models. Several authors, including \citet[][]{Munprecht2013, Mundprecht2015}, have embarked on the first 2D and 3D simulations of convection, which are essential for more realistic modeling of the convection–pulsation interaction in stars and provide valuable guidelines for improving the convective-turbulence treatment also in 1D hydrodynamical codes.

The paper is organized as follows: Section 2 provides an overview of the updated model datasets and discusses the adopted opacity tables, along with their effects on linear pulsation models for both CC and RRL stars. Section 3 investigates the impact of the adopted opacity tables on nonlinear pulsation models, including the topology of the instability strip and the shape and amplitude of bolometric light curves. It also analyzes the effect on inferred model parameters such as periods, mean magnitudes, and amplitudes, derived by transferring model predictions into the Johnson-Cousins photometric system, and explores how the Period-Wesenheit (PW) relations for CCs are affected by these updated opacity tabulations. Final remarks and conclusions close the paper.

\section{Radiative Rosseland opacity and pulsation code}

The calculation of pulsation models using the Stellingwerf code \citep[see][for details] {BonoStel1994, BonoCastMarc1999, Stellingwerf1982} is carried out in two steps. First, a linear hydrodynamic code is used to compute the stability of linear nonadiabatic pulsation models while neglecting convection.

The static envelope structure provided by the linear nonadiabatic code is then perturbed with a surface velocity modulated within the envelope layers, according to the radial eigenfunction corresponding to each selected mode. The full-amplitude pulsation behavior is then investigated using a nonlinear hydrodynamic code, which includes a non-local and time-dependent treatment of convection and its coupling with pulsation. In this approach, a mixing length parameter is adopted only to close the nonlinear system of hydrodynamical and convective equations. For details on the adopted convective model, the interested reader is referred to \citet{BonoStel1994, BonoCastMarc1999}.

Once a stable full-amplitude limit cycle is reached, the nonlinear code provides accurate predictions for the variation of all the stellar properties, such as luminosity, radius, radial velocity, temperature, and gravity, across the pulsation cycle. Additionally, the investigation of the pulsation behavior at the stable limit cycle enables predictions of the blue and red edges of the IS for each selected pulsation mode and any given assumptions about the mass, luminosity, and chemical composition of the pulsator.

This study investigates the effect of updating the opacity tables in both the linear and nonlinear pulsation codes on the predicted pulsation properties and stability of CC and RR Lyrae models.

In a previous study, \citet[][]{Bono_Incerpi1996} incorporated OPAL opacity tables \citep{Rogers_Iglesias1992, Iglesias_opal} and opacities by \citet[][]{Alexander1994} for temperatures lower than $10000K$ into our pulsation code, along with the traditional Stellingwerf Analytical Approximation, to analyze the effects of metallicity on RR Lyrae variables. By enhancing the resolution of opacity tables and accurately evaluating the derivatives, the authors ensured precise modeling of pulsational properties. 

Both the OPAL and low-T opacities implemented in the code by \citet[][]{Bono_Incerpi1996} were based on the solar heavy element distribution provided by \cite{Grevesse1991}.

The pulsation models by \citet[][hereinafter DS20, DS22]{Desomma2020, Desomma2022} for CCs and by \citet[][hereinafter M15]{Marconi2015} for RRLs were calculated using an updated version of the OPAL radiative opacities \citep{Iglesias_Rogers1996} and the molecular opacities by \citet[][]{Alexander1994}. These models will be compared with the models computed in the present work.

The sources for the radiative Rosseland opacity adopted in the present analysis are the same as those adopted in the evolutionary stellar model computations provided by the BaSTI\footnote{The BaSTI stellar model repository can be found at: http://basti-iac.oa-abruzzo.inaf.it} group \citet[][]{Hidalgo2018}. Specifically, opacities are from the latest release of the OPAL calculations \cite{Iglesias_Rogers1996} for temperatures greater than log(T) = 4.0, while calculations by \citet[][]{Ferguson_Alexander2005} - which include contributions from molecules and grains – have been adopted for lower temperatures. Both high- and low-temperature opacity tables have been computed for the solar-scaled heavy element distribution provided by \citet[][]{Caffau2011}, supplemented where necessary by the abundances given by\citet[][]{Lodders2010}.

To investigate the impact of updating the radiative opacity tables on CC and RRL pulsation predictions, we computed models covering different masses, luminosities, and chemical composition exploring in detail the range of effective temperatures where these stars are expected to be pulsationally unstable. From these computations, we derived pulsation observables such as periods, the topology of the IS, and light curves, along with the corresponding amplitudes, mean magnitudes, and colors. We then compared these results with those based on the same pulsation code but using older opacity tabulations (see the previous discussion).

For the CC computations, the luminosity for each model was fixed according to the mass-luminosity (ML) relation predicted by the evolutionary models presented by \citet{BonoTorn2000}. Specifically, we adopted the luminosity levels obtained by increasing, for each fixed stellar mass, the luminosity predicted by canonical stellar models, i.e. those not accounting for core convective overshooting during the central H-burning phase (we refer the reader to the quoted paper and DS20 for more details) by $\Delta\log(L/L_\odot)=0.2$ dex. In the case of RRLs, the faintest luminosity value was selected for each given metallicity, corresponding to the Zero-Age Horizontal Branch (ZAHB) luminosity \citepalias[see][for details]{Marconi2015}. For both variables, a unique mixing length value, $\alpha_{ml}$ = $l/H_{P}$ =1.5 \footnote{$l$ is the length of the path covered by the convective elements before releasing heat, and $H_{P}$ is the local pressure scale height.}, was adopted to close the system of nonlinear equations.

Two distinct chemical compositions were selected for each class of pulsating stars: Z=0.004; Y=0.25 and Z=0.02; Y=0.28 for CCs and Z=0.001; Y=0.245 and Z=0.02; Y=0.28 for RRLs. We assessed the modal stability of both the fundamental (F) and first-overtone (FO) pulsation modes for the various combinations of mass, luminosity and effective temperature.  More details about the model calculation can be found in the aforementioned papers. 

Tables \ref{f_fo_param_model_cc} and \ref{f_fo_param_model_rrl} present the parameters for the computed pulsation models for CCs and RRLs, respectively. The columns in these tables report the chemical abundances (Z and Y), the pulsation mode, the adopted stellar mass (in solar units), luminosity (in solar units), effective temperature (in Kelvin), the obtained period (in days), and logarithmic radius (in solar units).

\begin{table*}
\centering
\caption{\label{f_fo_param_model_cc} Intrinsic stellar parameters for F and FO mode Classical Cepheid models for the selected chemical compositions. The whole table is accessible in a machine-readable form.}
\centering
\begin{tabular}{cccccccc}
\hline\hline
Z & Y & Pulsation Mode & \msun & \lsun & $T_{eff}$[K] & P[d] & $\log(\overline{R}$/$R_{\odot})$ \\
\hline\hline
&&& Classical Cepheids &&&\\
\hline
0.004 &  0.25 &  F &   4.0 &  3.11 &  5400 & 5.44902 &  1.611 \\
0.004 &  0.25 &  F &   4.0 &  3.11 &  5500 & 5.13436 &  1.599 \\
0.004 &  0.25 &  F &   4.0 &  3.11 &  5600 & 4.84289 &  1.586 \\
...\\
0.004 &  0.25 &  FO &  4.0 &  3.11 &  5900 & 2.91818 &  1.543 \\
0.004 &  0.25 &  FO &  4.0 &  3.11 &  6100 & 2.58571 &  1.514 \\
0.004 &  0.25 &  FO &  4.0 &  3.11 &  6200 & 2.45770 &  1.499 \\
...\\
0.02 &  0.28 &  F &   4.0 &  2.94 &  5300 & 4.32575 &  1.546 \\
0.02 &  0.28 &  F &   4.0 &  2.94 &  5400 & 4.06813 &  1.531 \\
0.02 &  0.28 &  F &   4.0 &  2.94 &  5500 & 3.82069 &  1.516 \\
...\\
\hline
\end{tabular}
\end{table*}

\begin{table*}
\centering
\caption{\label{f_fo_param_model_rrl} As Table~\ref{f_fo_param_model_cc} but for the computed RR Lyrae pulsation models.}
\centering
\begin{tabular}{ccccccccc}
\hline\hline
Z & Y & Pulsation Mode & \msun & \lsun & $T_{eff}$[K] & P[d] & $\log(\overline{R}$/$R_{\odot})$ \\
\hline\hline
&&&&RR Lyrae&&&&\\
\hline
0.001 & 0.245 & F & 0.58 &  1.87 &  5700 & 1.43834 &  0.935 \\
0.001 & 0.245 & F & 0.58 &  1.87 &  5800 & 1.36031 &  0.924 \\
0.001 & 0.245 & F & 0.58 &  1.87 &  5900 & 1.28939 &  0.915 \\
...\\
0.001 &  0.245 & FO &  0.58 &  1.87 &  6400 & 0.70827 &  0.856 \\
0.001 &  0.245 & FO &  0.58 &  1.87 &  6500 & 0.67410 &  0.843 \\
0.001 &  0.245 & FO &  0.58 &  1.87 &  6600 & 0.64034 &  0.829 \\
...\\
0.02 &  0.28 & F &  0.51 &  1.69 &  5600 & 1.23075 &  0.870 \\
0.02 &  0.28 & F &  0.51 &  1.69 &  5700 & 1.16263 &  0.857 \\
0.02 &  0.28 & F &  0.51 &  1.69 &  5800 & 1.09190 &  0.842 \\
\hline
\end{tabular}
\end{table*}

\begin{figure*}
\centering
    \vbox{
    \hbox{
    \includegraphics[width=0.35\textwidth]{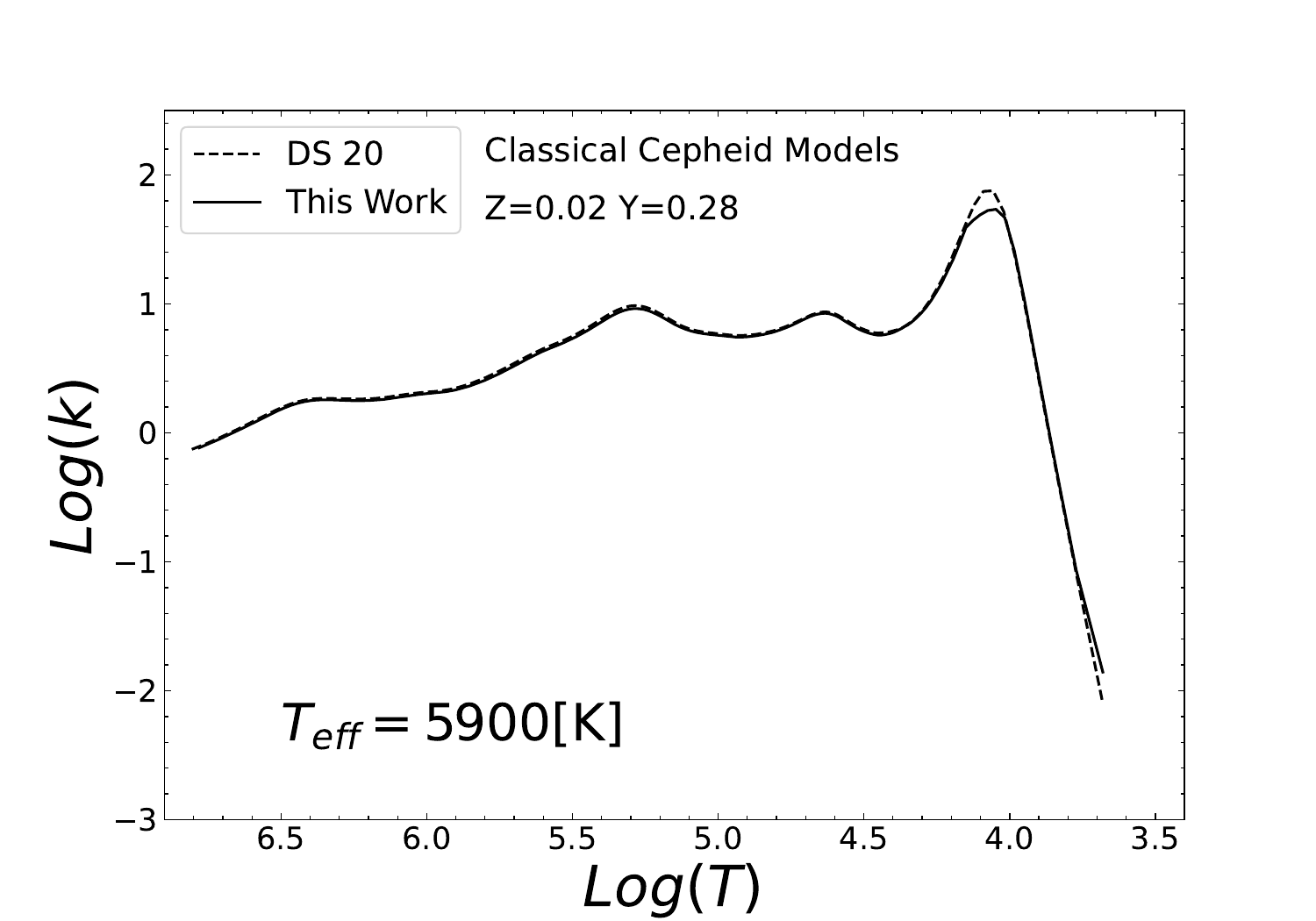}
    \includegraphics[width=0.35\textwidth]{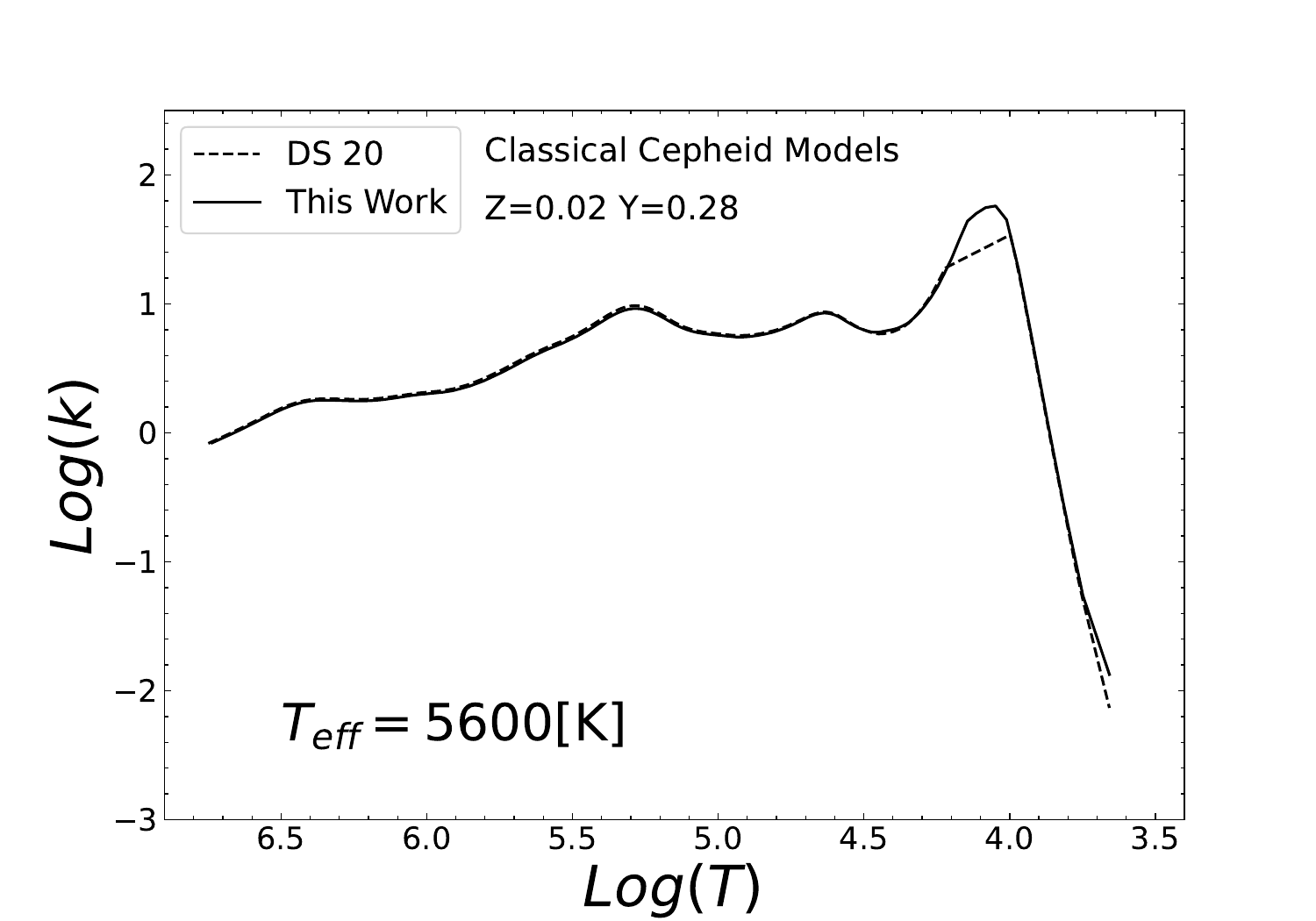}
    \includegraphics[width=0.35\textwidth]{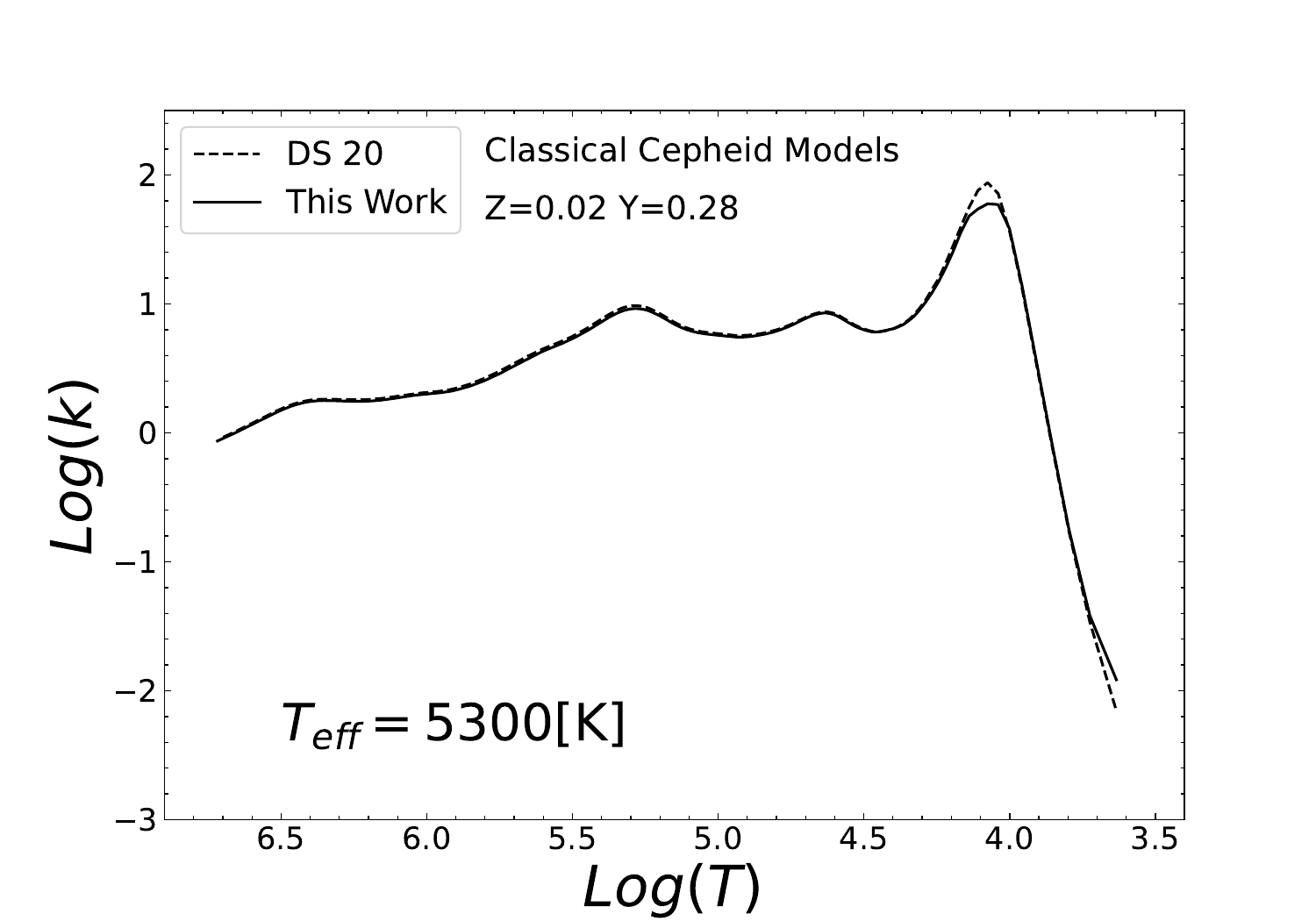}
    }
    \hbox{
    \includegraphics[width=0.35\textwidth]{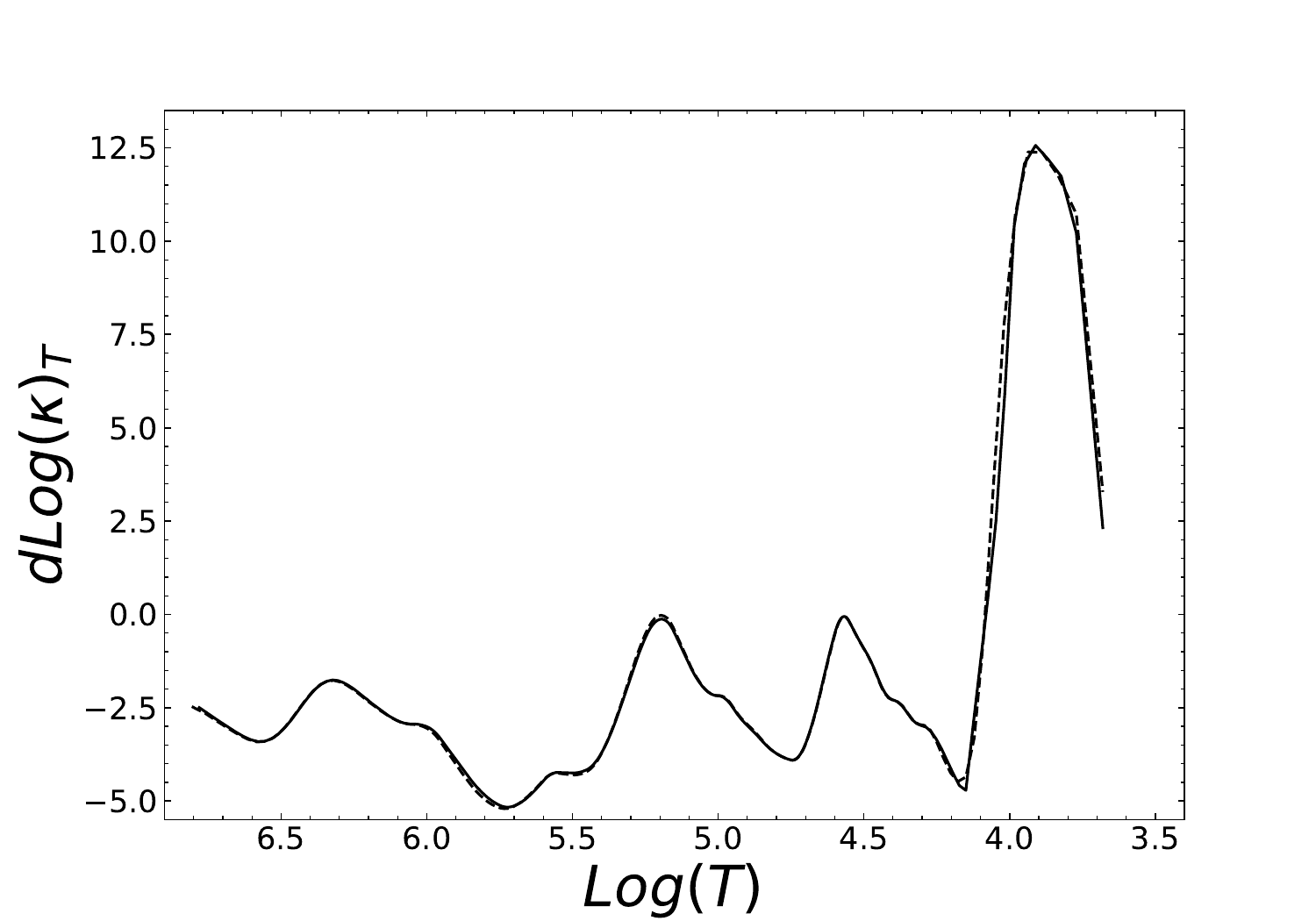}
    \includegraphics[width=0.35\textwidth]{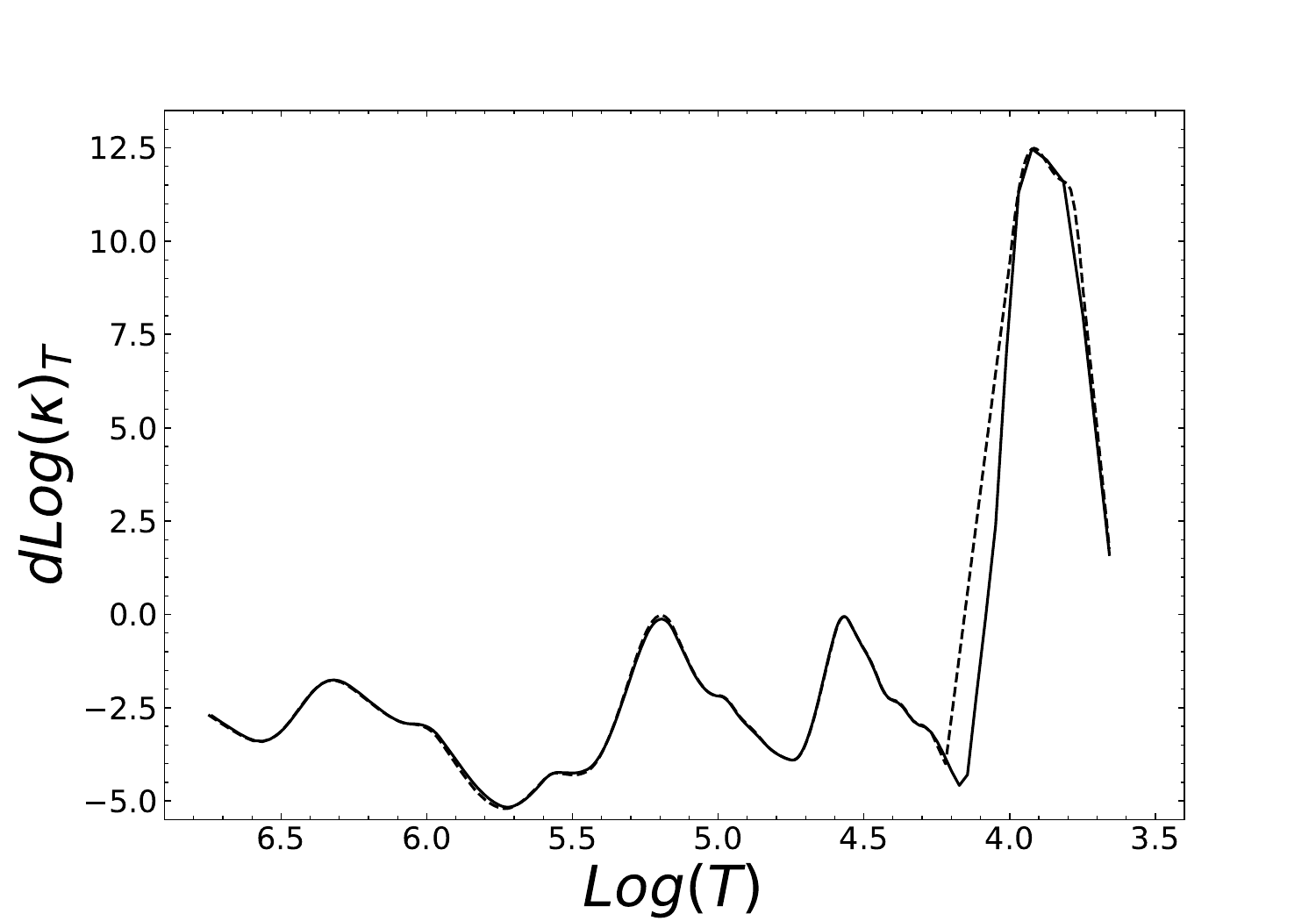}
    \includegraphics[width=0.35\textwidth]{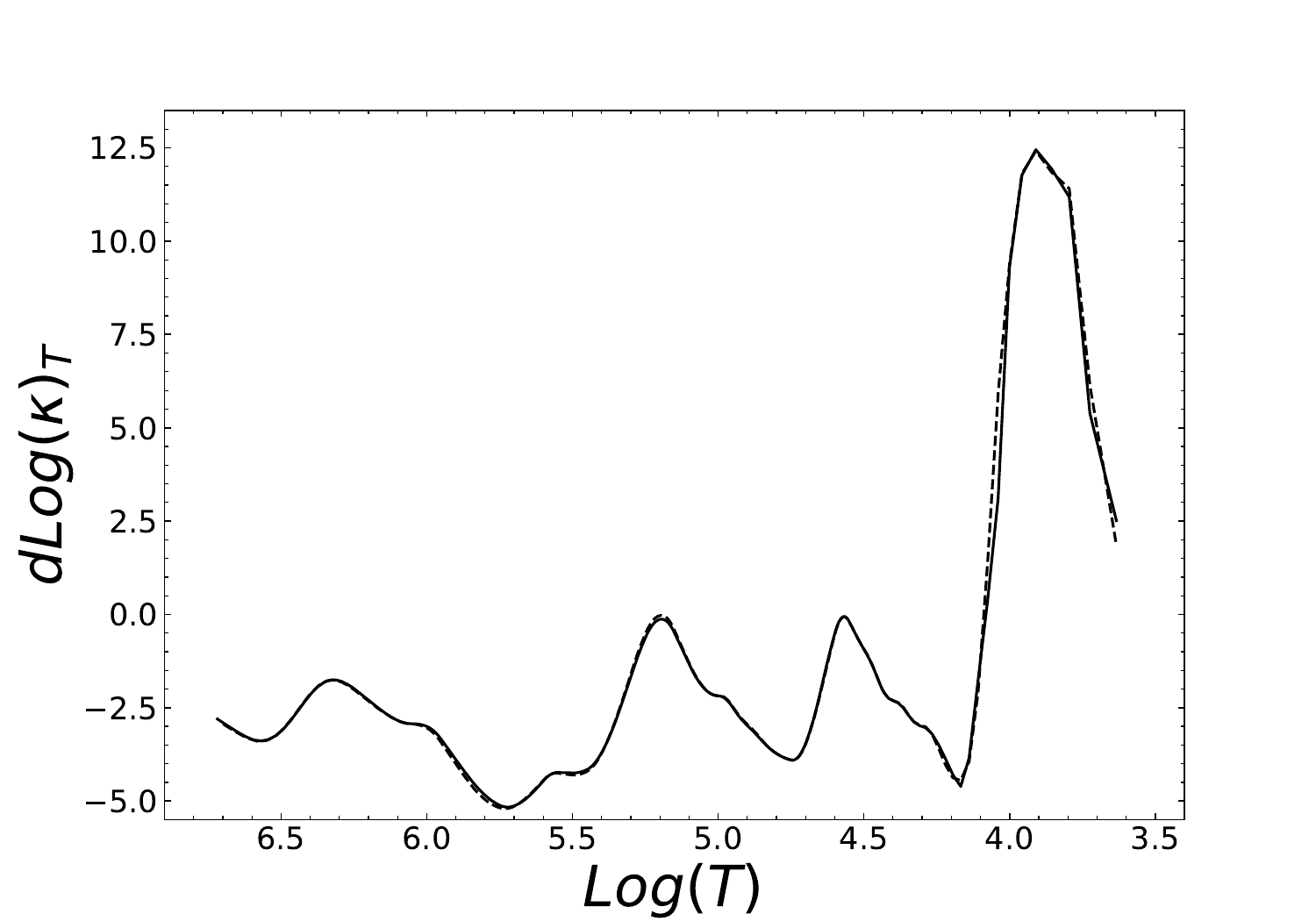}
    }
    \hbox{
    \includegraphics[width=0.35\textwidth]{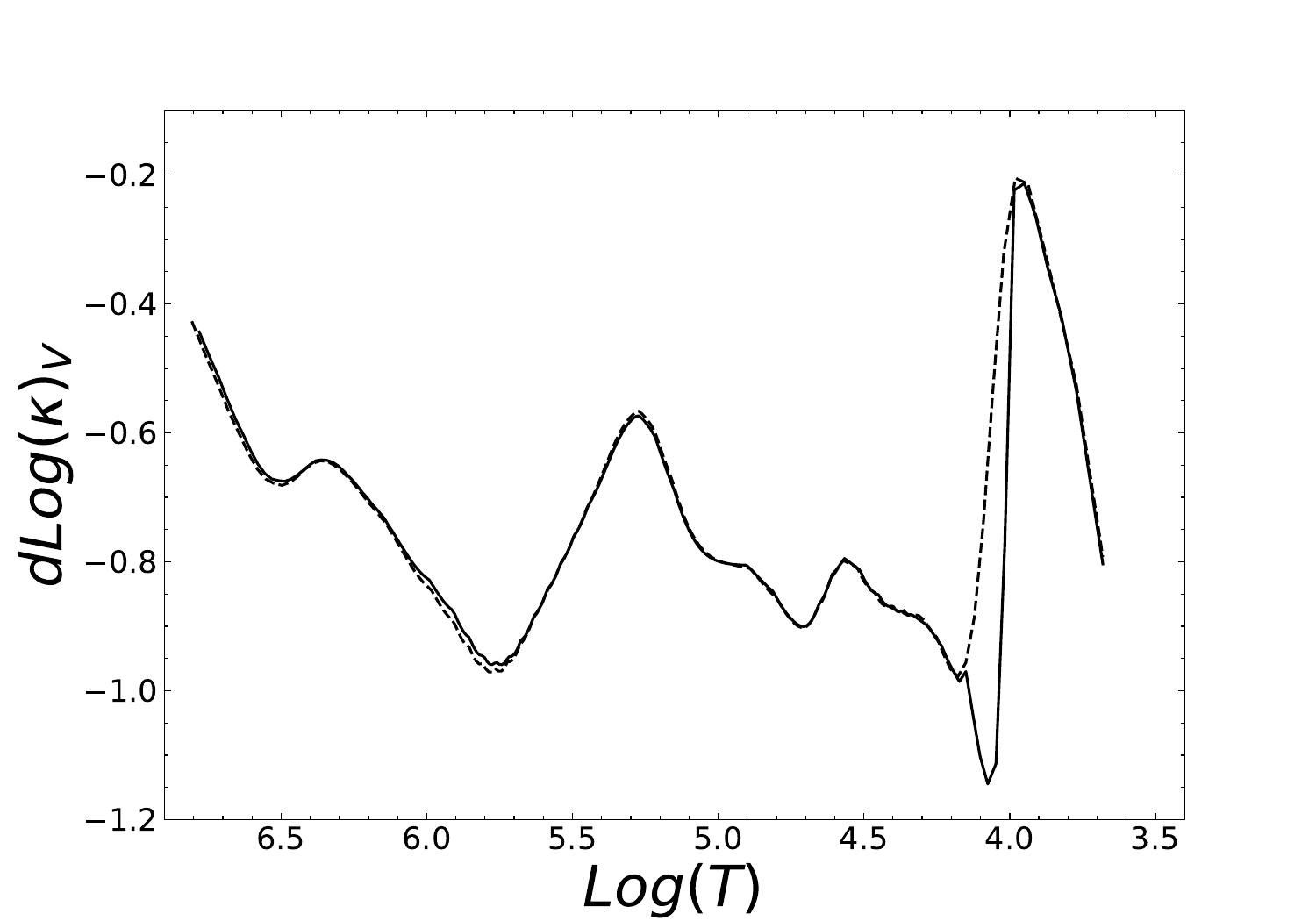}
    \includegraphics[width=0.35\textwidth]{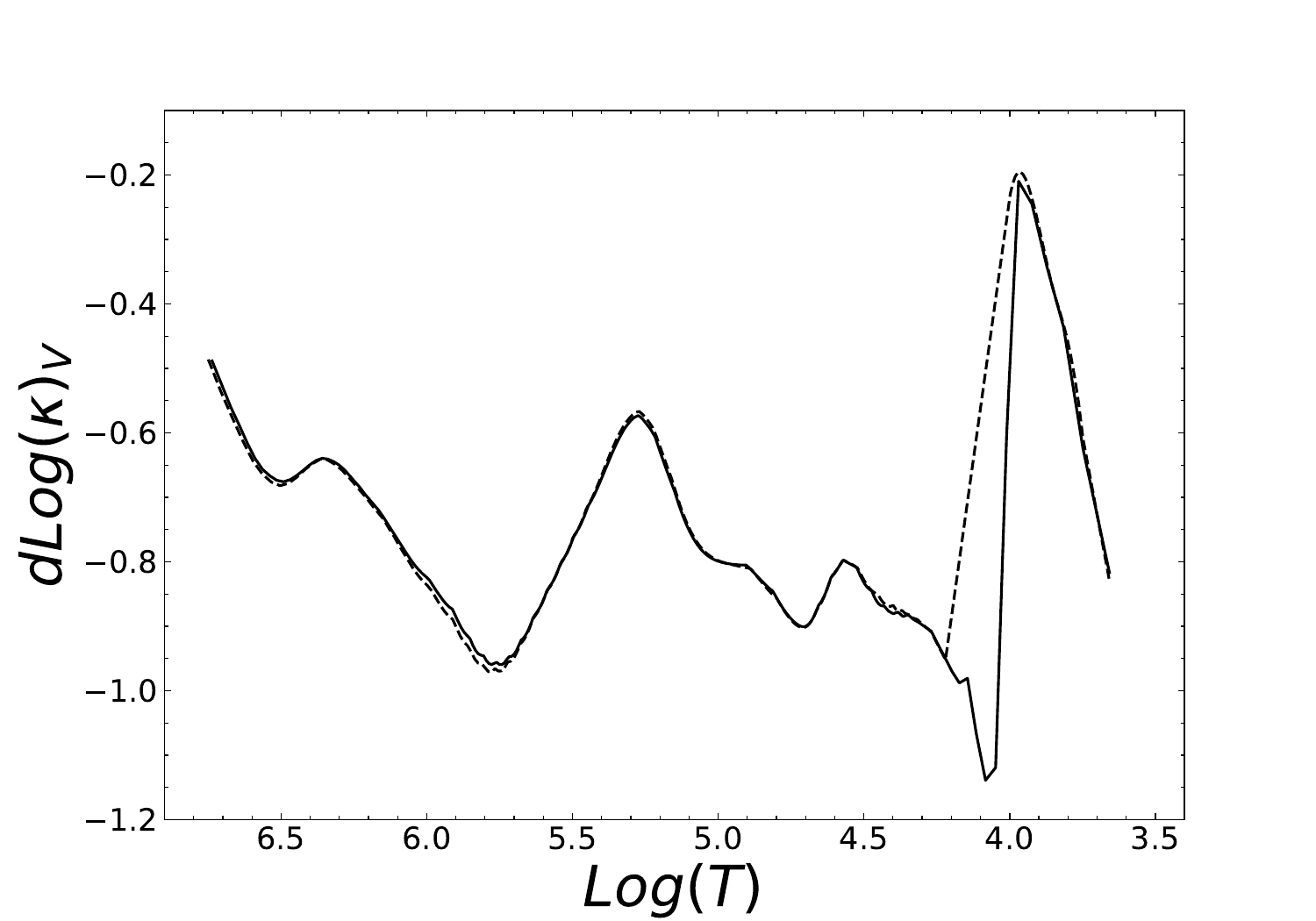}
    \includegraphics[width=0.35\textwidth]{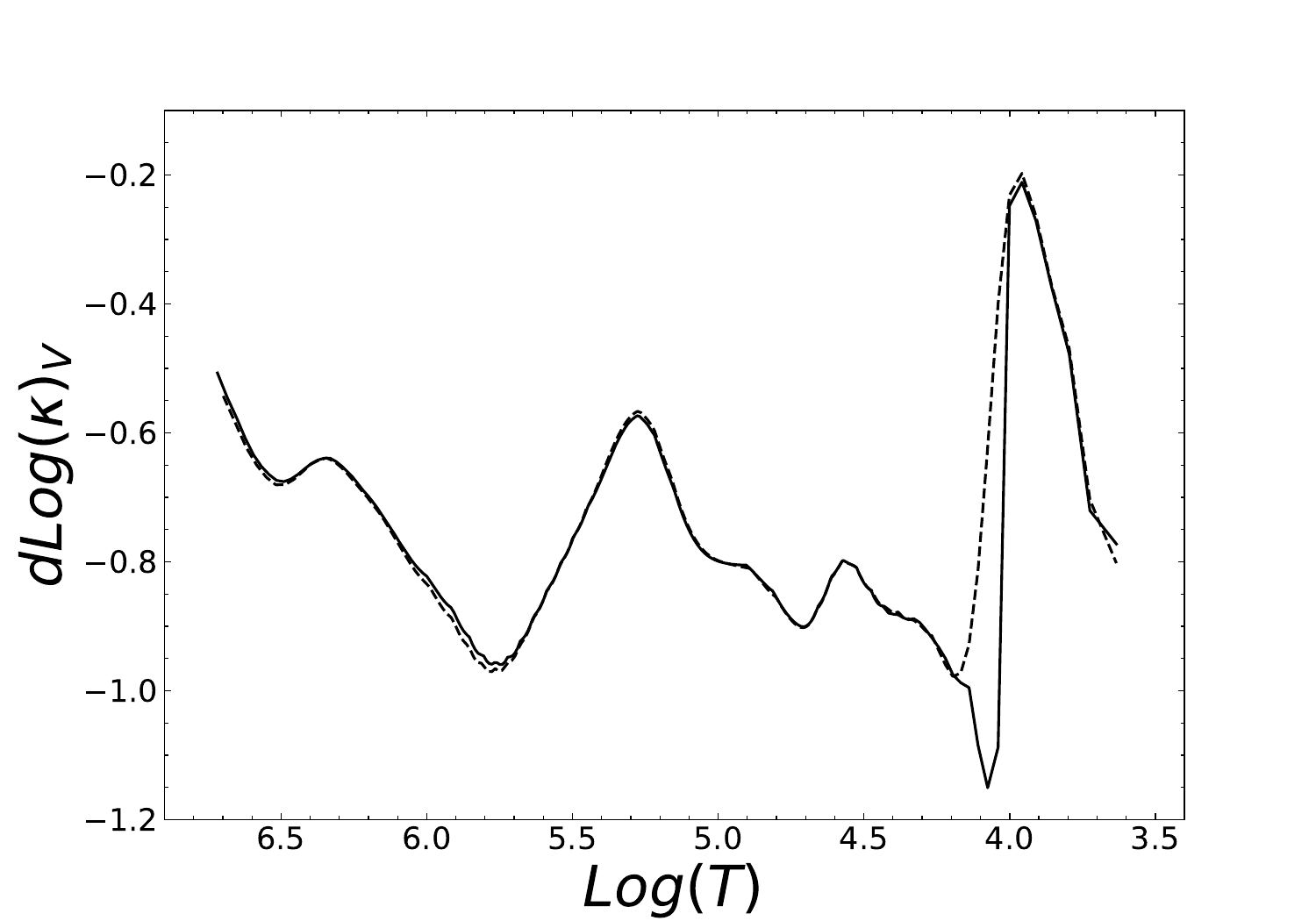}
    }}
    \caption{\label{fig:op_mw_cc} The trend of the logarithmic opacity (top panels), the derivatives of opacity with respect to temperature (middle panels), and with respect to volume (bottom panels) derived from linear and radiative models of a $4M_\odot$ CC with a luminosity of $\log(L/L_\odot)=2.94$, $Z = 0.02$ and $Y = 0.28$, as obtained using various opacity sets. The dashed lines represent results from the models presented in \citetalias{Desomma2022}, while the solid lines represent results from the present computations. The left panels show the results corresponding to a pulsation model near the blue edge of the instability strip (IS) in the H-R diagram, the central panels correspond to a model at the midpoint of the IS temperature range, and the right panels show the behavior near the red edge of the IS (see labels)).}
\end{figure*}

Figure \ref{fig:op_mw_cc} shows the trend of opacity, the temperature derivatives ($(dlog\kappa)_T$ = $dlog\kappa/dlogT$) and the volume derivatives ($(dlog\kappa)_V$ = $dlog\kappa/dlogV$) of opacity as a function of temperature, as derived from linear pulsation models of CCs for the adopted solar chemical composition. The figure presents the behavior of these physical quantities from the current computations and from those performed by \citetalias{Desomma2022}, using the same assumptions about stellar mass, luminosity, and effective temperature. For the selected luminosity, three distinct $T_{eff}$ values have been chosen, corresponding to
a position close to the blue boundary, the middle region, and close to the red boundary of the IS, respectively.

\begin{figure*}
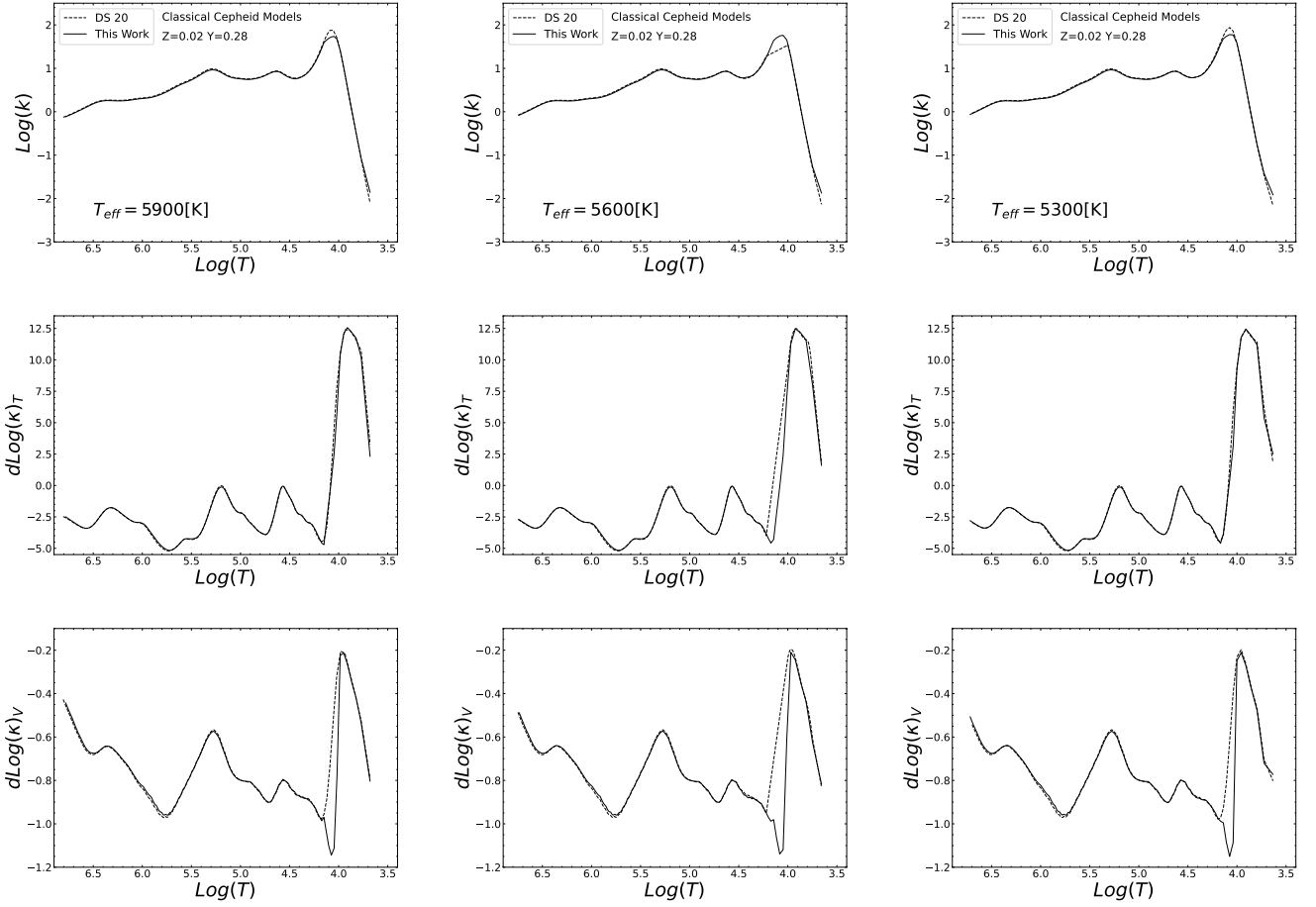

\centering
    \vbox{
    \hbox{
    \includegraphics[width=0.33\textwidth]{figures/plot_logk_5900_2p94_4p0_B_z22y28.pdf}
    \includegraphics[width=0.33\textwidth]{figures/plot_logk_5600_2p94_4p0_B_z22y28.pdf}
    \includegraphics[width=0.33\textwidth]{figures/plot_logk_5300_2p94_4p0_B_z22y28.pdf}
    }
    \hbox{
    \includegraphics[width=0.33\textwidth]{figures/plot_dk_dT_5900_2p94_4p0_B_z22y28.pdf}
    \includegraphics[width=0.33\textwidth]{figures/plot_dk_dT_5600_2p94_4p0_B_z22y28.pdf}
    \includegraphics[width=0.33\textwidth]{figures/plot_dk_dT_5300_2p94_4p0_B_z22y28.pdf}
    }
    \hbox{
    \includegraphics[width=0.33\textwidth]{figures/plot_dk_dV_5900_2p94_4p0_B_z22y28.pdf}
    \includegraphics[width=0.33\textwidth]{figures/plot_dk_dV_5600_2p94_4p0_B_z22y28.pdf}
    \includegraphics[width=0.33\textwidth]{figures/plot_dk_dV_5300_2p94_4p0_B_z22y28.pdf}
    }}
    \caption{\label{fig:op_smc_cc} As in Fig~\ref{fig:op_mw_cc}, but for CC pulsation models for the chemical composition: $Z=0.004$ $Y=0.25$.}
\end{figure*}

From the data shown in this figure, one can notice that the differences in the opacity profiles of the two pulsational models, computed under different assumptions regarding radiative opacities, are quite small. On average, the differences are below or of the order of 5\% along the whole stellar envelope, reaching a maximum value of the order of 10\% at the location of the opacity peak at $\log{T}\sim 4.0$.
Consequently, it is not surprising that a similar pattern is found in the derivatives of opacity with respect to temperature and volume. The most significant differences occur in the correspondence of the temperature regime near the opacity profile peak, around the H ionization region. A similar trend, shown in Figure \ref{fig:op_smc_cc} has been observed for the CC models at metallicity Z=0.004. 

\begin{figure*}
\centering
    \vbox{
    \hbox{
    \includegraphics[width=0.33\textwidth]{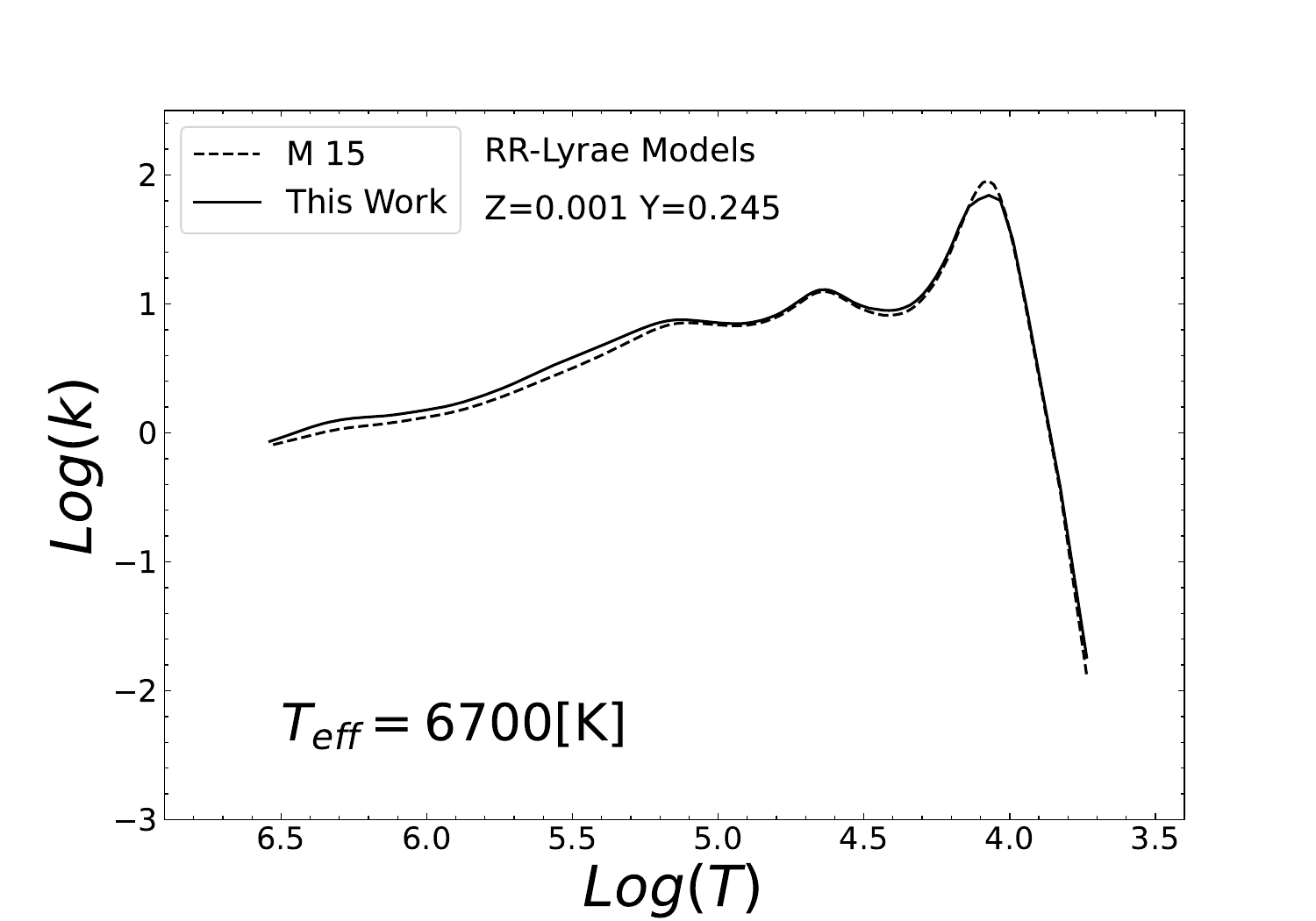}
    \includegraphics[width=0.33\textwidth]{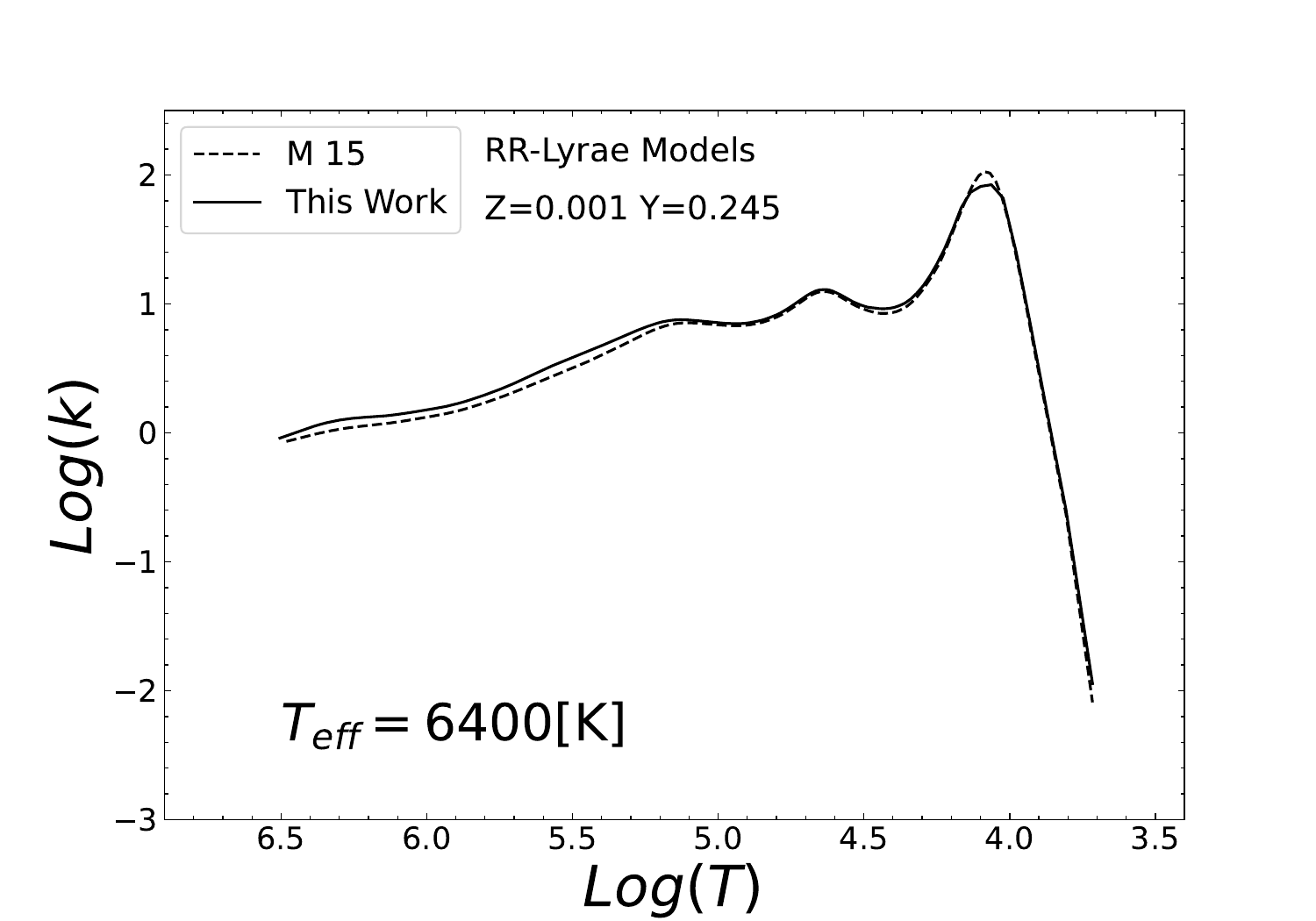}
    \includegraphics[width=0.33\textwidth]{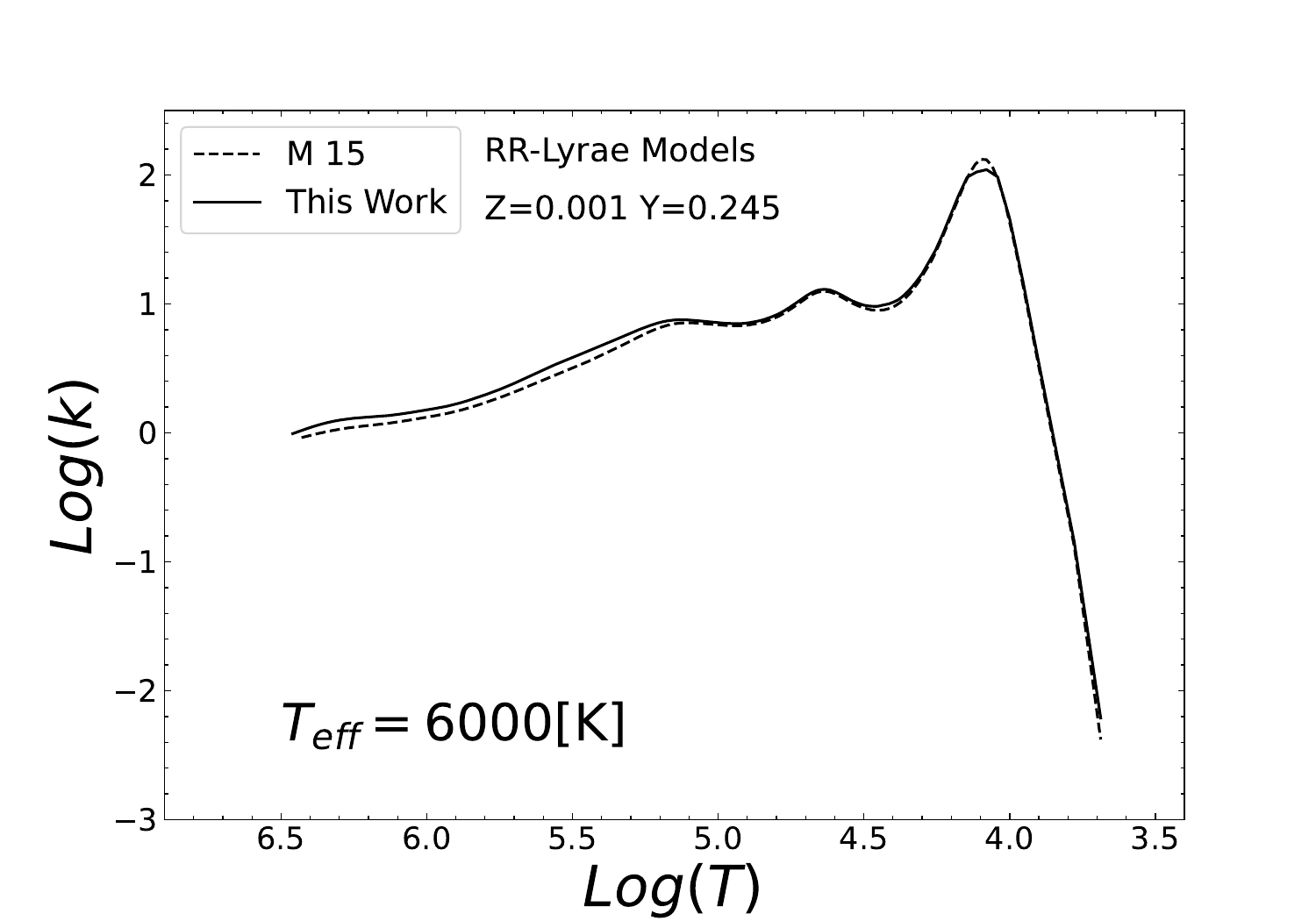}
    }
    \hbox{
    \includegraphics[width=0.33\textwidth]{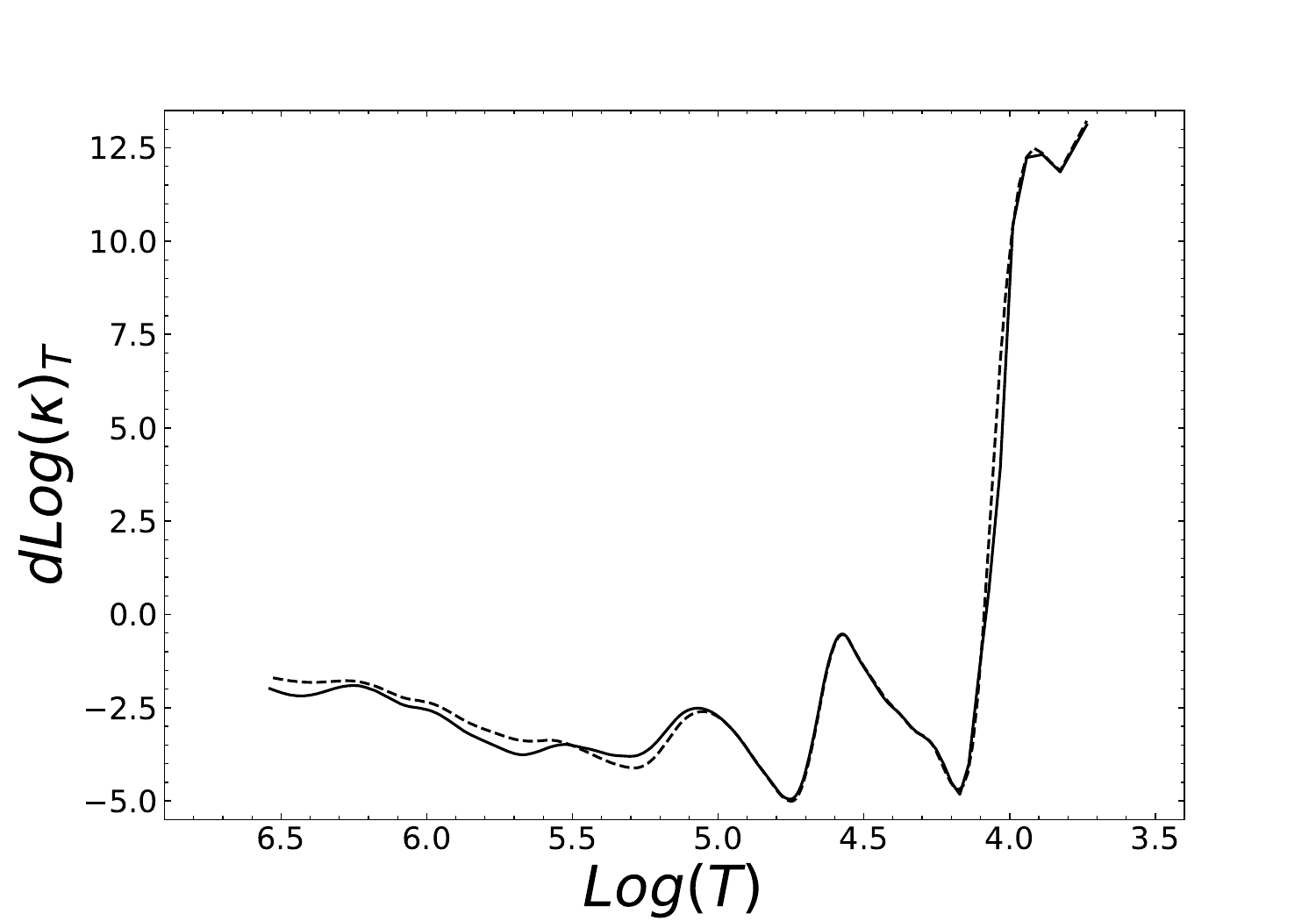}
    \includegraphics[width=0.33\textwidth]{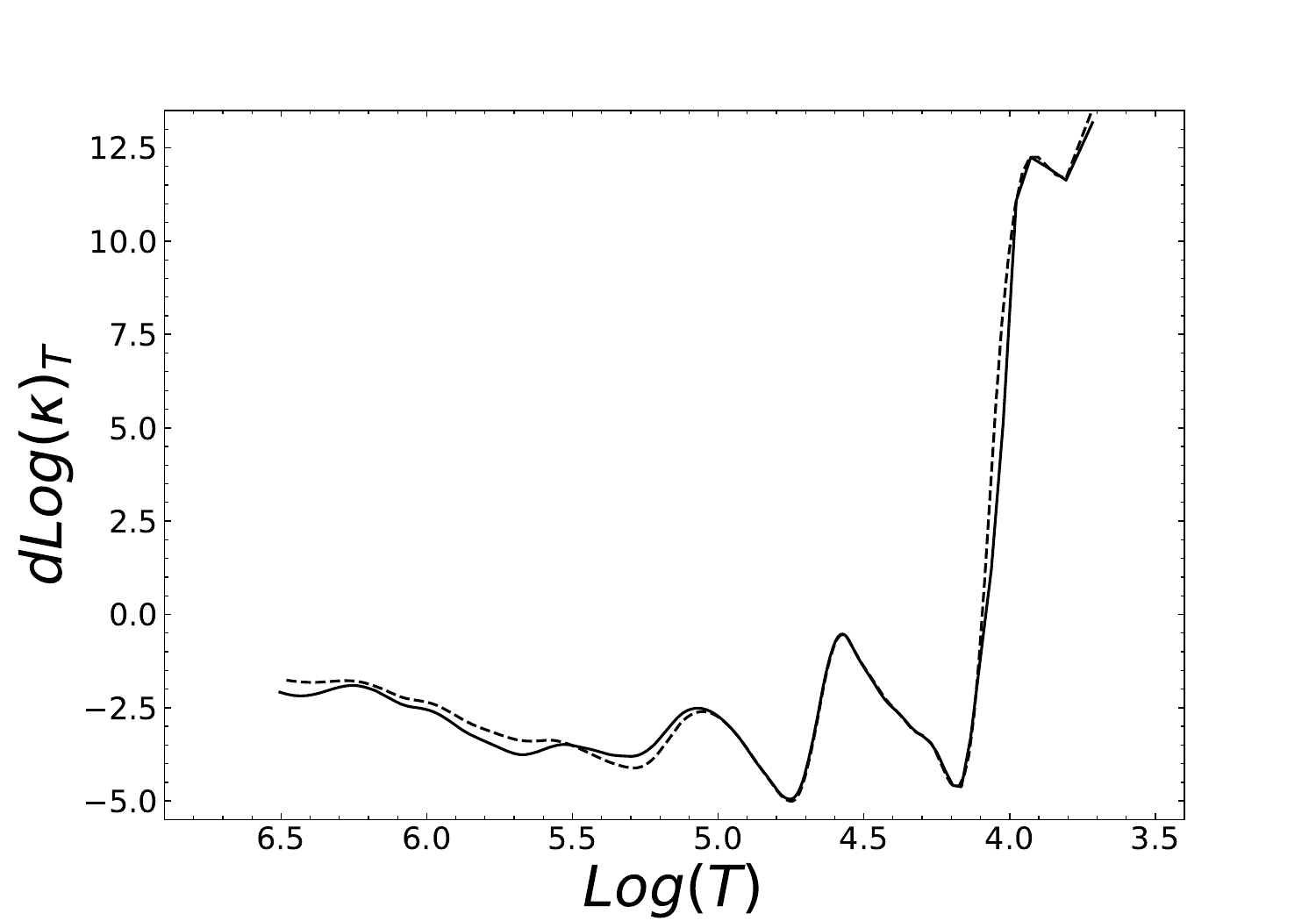}
    \includegraphics[width=0.33\textwidth]{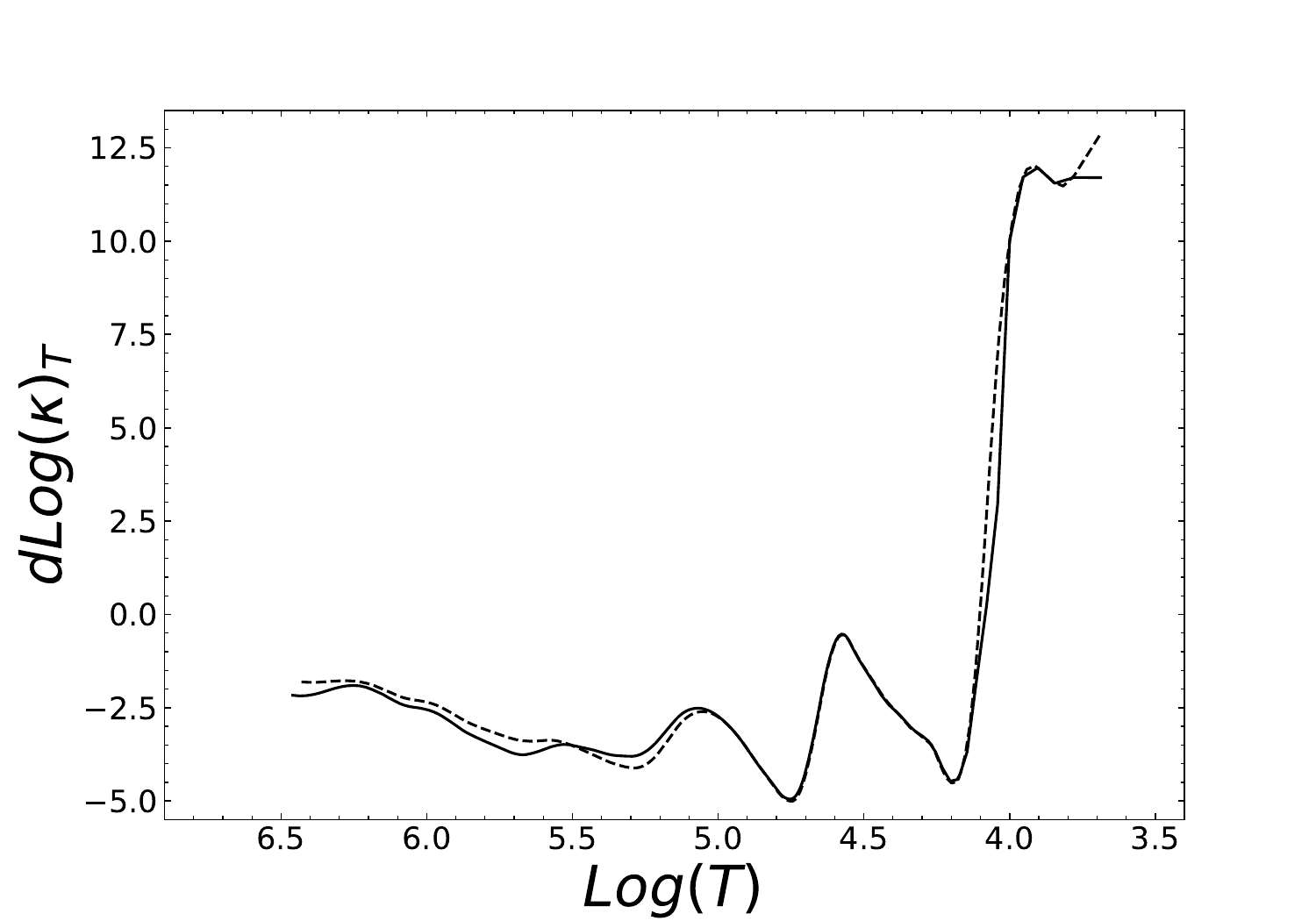}
    }
    \hbox{
    \includegraphics[width=0.33\textwidth]{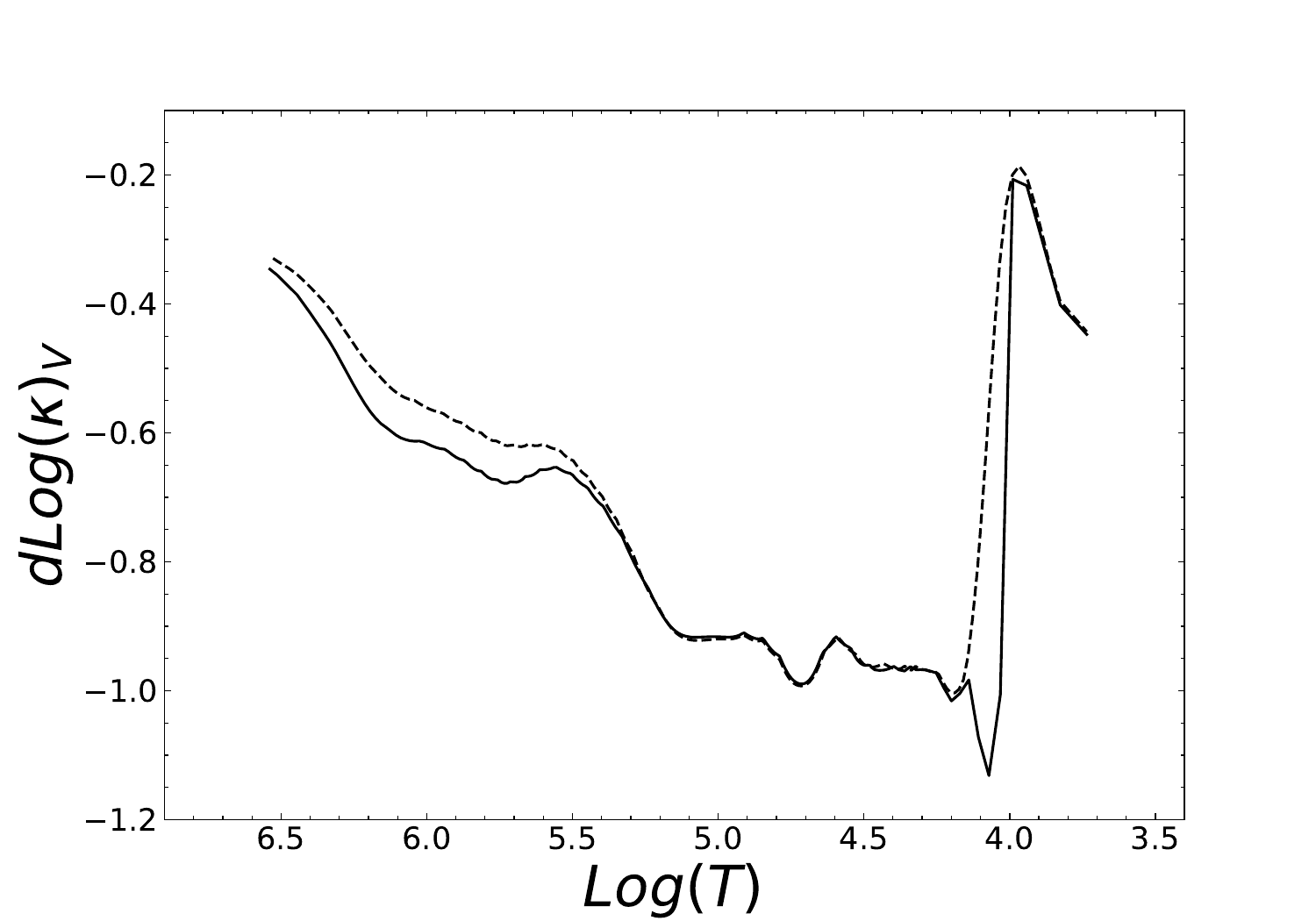}
    \includegraphics[width=0.33\textwidth]{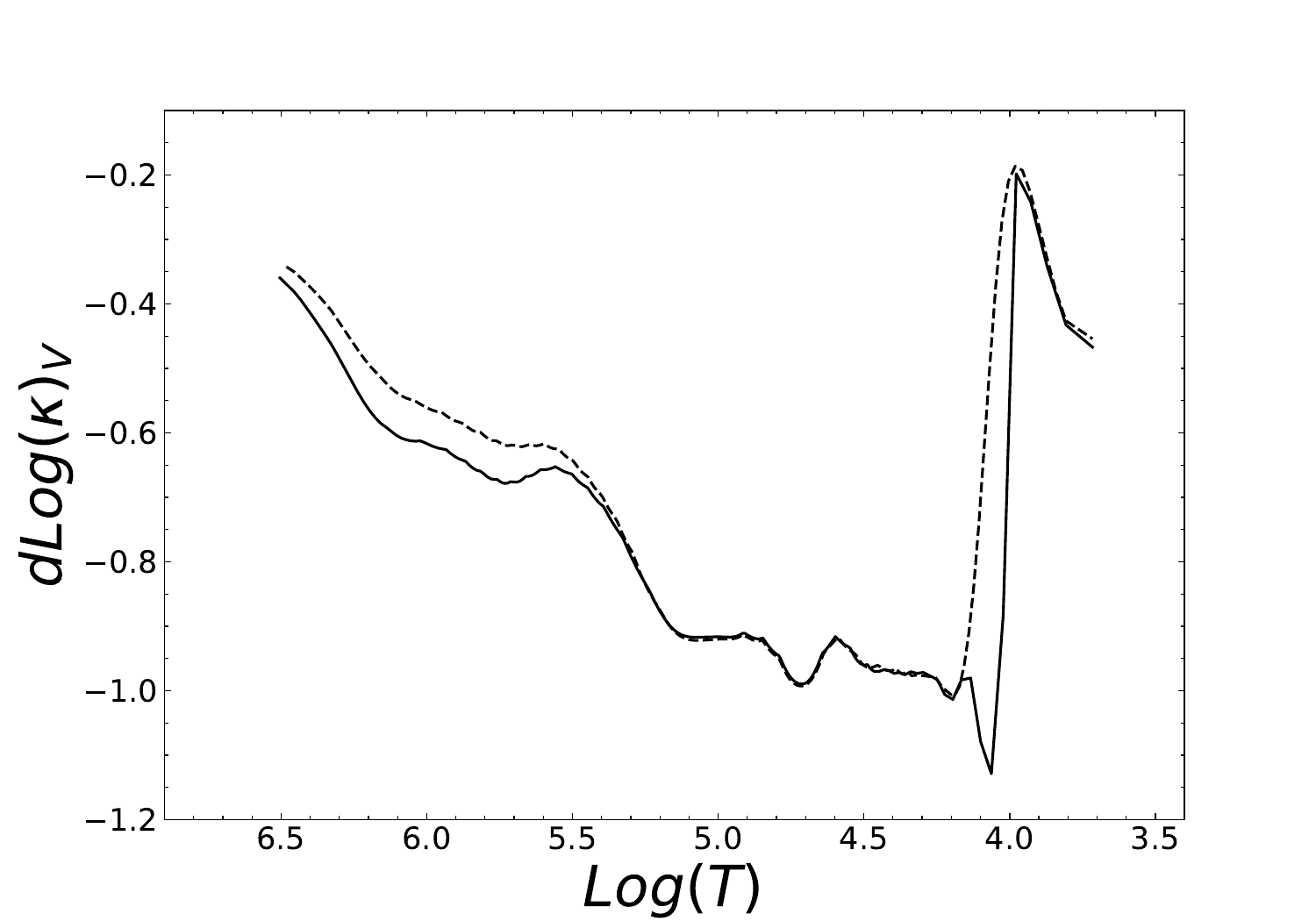}
    \includegraphics[width=0.33\textwidth]{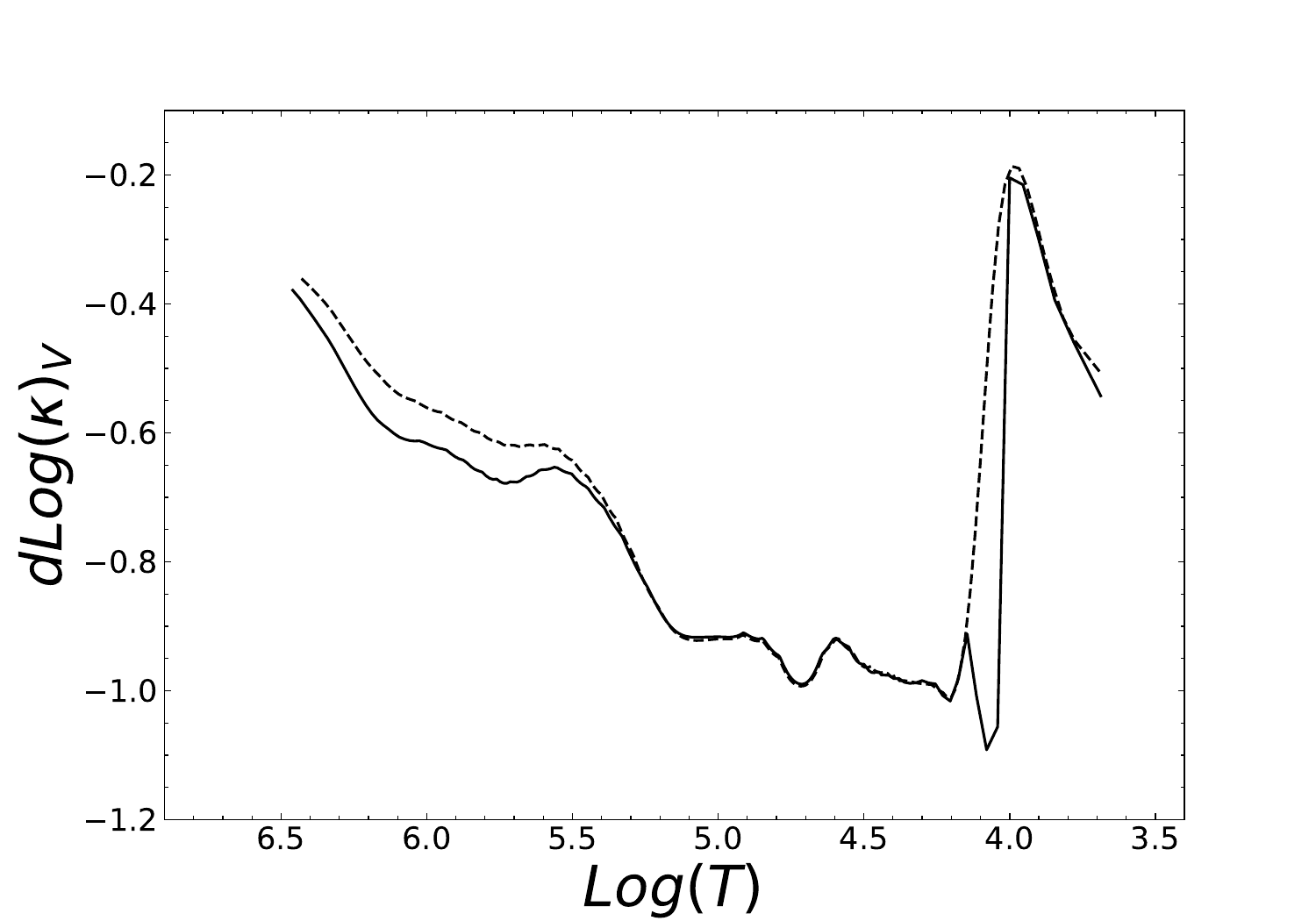}
    }}
    \caption{\label{fig:op_z13_rrl} As in Fig~\ref{fig:op_mw_cc}, but for RR Lyrae pulsation models for the chemical composition: $Z=0.001$ $Y=0.245$.}
\end{figure*}

Figure \ref{fig:op_z13_rrl} shows the same comparison as Fig.~\ref{fig:op_mw_cc}, but for RR Lyrae pulsation models with $Z=0.001$. In this case, the differences in the opacity profile are slightly larger than those for CC, being around $\sim15-20$\% for temperatures below $\log{T}\approx5.0$ and at the peak of the opacity profile at $\log{T}\approx4.0$. This impacts, the opacity derivatives, particularly the derivative with respect to volume at constant temperature. A similar behavior is found for the solar metallicity RRL models.

Previous comparisons reveal that the change of the adopted radiative opacity tabulations does not lead to significant variations in the predictions from linear, radiative pulsation computations."

%\begin{figure*}[h]

\clearpage

%\begin{figure*}[h]

\section{Updated opacity tables and non-linear pulsation models}

This section analyzes the impact of using updated opacity tables on pulsation non-linear model predictions. Specifically, we discuss the effects on the topology of the IS, bolometric light curves, and periods.

\subsection{The topology of the instability strip}

The present theoretical framework enables us to predict the approach to nonlinear limit cycle stability of various pulsation modes, allowing us to constrain the IS topology, that is the region in the H-R diagram where radial modes reach a stable nonlinear limit cycle. A radial mode is considered to approach a stable nonlinear limit cycle when period and amplitudes over consecutive cycles achieve their asymptotic behavior. 

To determine the boundaries of the IS, we adopted a step size of 100 K in effective temperature for each given chemical composition and luminosity level.  The temperature sampling becomes slightly coarser across the IS. The effective temperatures of the IS edges for F and FO CC and RRL pulsators, along with the adopted chemical compositions, are listed in Table~\ref{boundaries}. The blue (red) edges are defined as 50 K hotter (cooler) than the first (last) pulsating model in the specific pulsation mode.

\begin{figure*}[ht]
    \centering
    \vbox{
    \hbox{
    \includegraphics[width=1.0\linewidth]{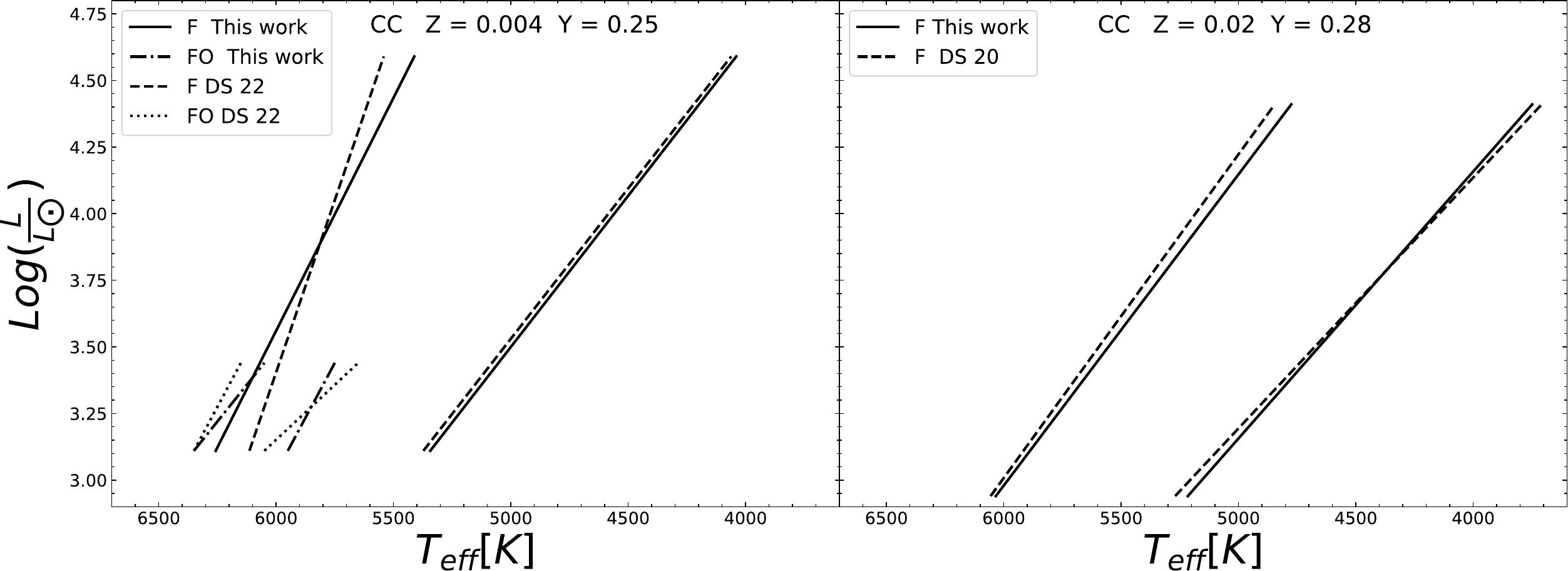}
    }
    \hbox{
    \includegraphics[width=1.0\textwidth]{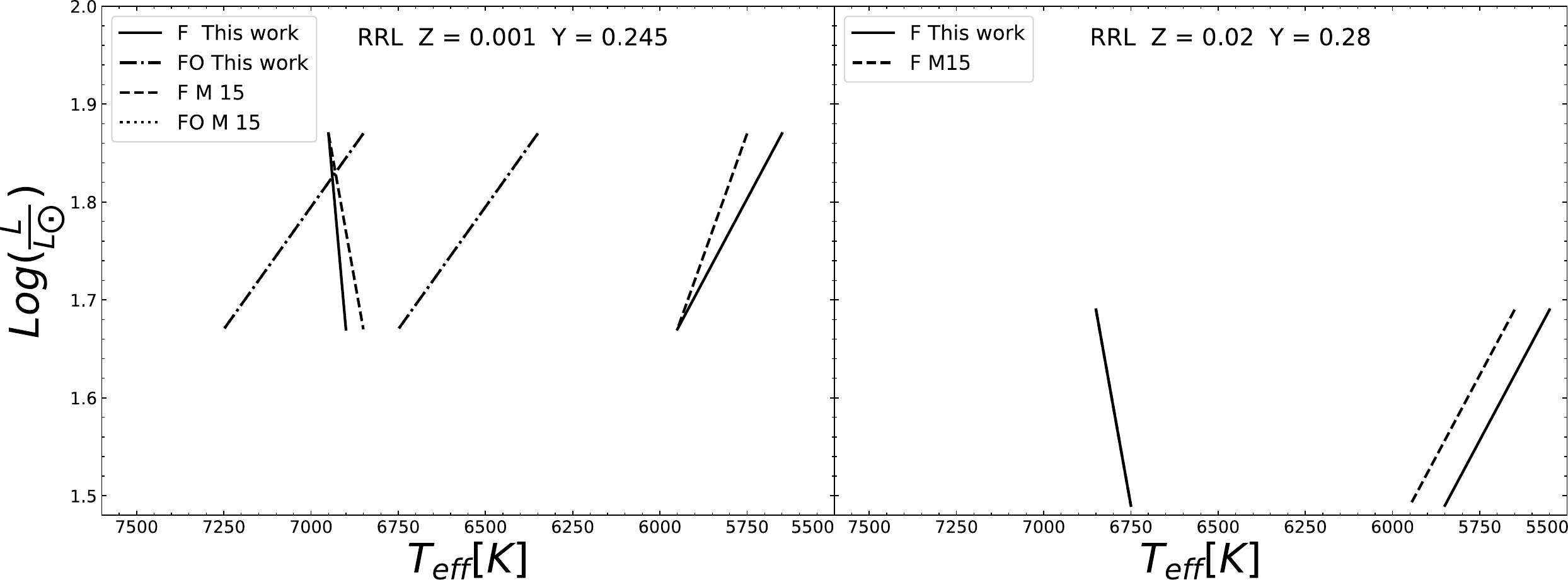}
    }}
    \caption{\label{fig:strips} Upper panels: The CC F and FO instability strips for $Z = 0.004$, $Y = 0.25$ (left panel) and $Z = 0.02$, $Y = 0.28$ (right panel). Solid and long-dashed lines represent the results of the present work for F and FO pulsators, respectively, while short-dashed and dotted lines represent the IS results for F and FO pulsators, respectively, from \citetalias{Desomma2020} and \citetalias{Desomma2022}. Bottom panels: Same as the upper panels but for RRL pulsators with $Z = 0.001$, $Y = 0.245$ (left panel) and $Z = 0.02$, $Y = 0.28$ (right panel). Solid and long-dashed lines represent the results of the present work for F and FO pulsators, respectively; while short-dashed and dotted lines represent results for F and FO pulsators, respectively, from \citetalias{Marconi2015}.
    Please note that different scales are used in the upper and lower panels due to the substantial difference in the temperature ranges covered by CCs and RR Lyrae stars.}
\end{figure*}

Figure \ref{fig:strips} shows the predicted instability strips for CCs(upper panels) and RRL (bottom panels) compared to those derived by \citetalias[][]{Desomma2020, Desomma2022} and by \citetalias[][]{Marconi2015}, respectively. The newly computed IS boundaries are in very good agreement, within $\pm$100~K.

Interestingly, the present computations provide plain support to the previously discussed trend of the IS getting redder as metallicity increases, both for CCs (\citetalias[][]{Desomma2022}, \citealt[][]{DeSomma2021}) and RRLs \citepalias{Marconi2015}, which has relevant implications for the behavior of the period-luminosity (PL) relations \citep[][]{Marconi2005}.

\begin{table*}
\centering
\caption{\label{boundaries} Predicted IS boundaries: first overtone blue edge (FOBE), fundamental blue edge (FBE), first overtone red edge (FORE), and fundamental red edge (FRE), for Classical Cepheids and RR Lyrae stars, for the labeled assumptions about the chemical composition. Due to the $T_{eff}$ step adopted for our model grid, the uncertainty in the location of the IS boundaries $\Delta{T_{eff}}=\pm\; 50\; K$.}
\centering
\begin{tabular}{cccccccc}
\hline\hline
Z & Y & \msun & \lsun & FOBE & FBE & FORE & FRE \\
\hline\hline
&&&Classical Cepheids &&&&\\
\hline
0.004 &  0.25 & 4.0 & 3.11 & 6350 & 6150 & 5850 & 5350 \\
0.004 &  0.25 & 5.0 & 3.44 & 6050 & 6050 & 5850 & 5050 \\
0.004 &  0.25 & 6.0 & 3.7 &  & 5950 &  & 4850 \\
0.004 &  0.25 & 7.0 & 3.93 &  & 5950 &  & 4550 \\
0.004 &  0.25 & 8.0 & 4.12 &  & 5650 &  & 4250 \\
0.004 &  0.25 & 9.0 & 4.29 &  & 5550 &  & 4250 \\
0.004 &  0.25 & 10.0 & 4.45 &  & 5450 &  & 4250 \\
0.004 &  0.25 & 11.0 & 4.59 &  & 5450 &  & 4250 \\
0.02 &  0.28 & 4.0 & 2.94 &  & 6050 &  & 5250 \\
0.02 &  0.28 & 5.0 & 3.27 &  & 5750 &  & 4850 \\
0.02 &  0.28 & 6.0 & 3.53 &  & 5550 &  & 4650 \\
0.02 &  0.28 & 7.0 & 3.76 &  & 5250 &  & 4350 \\
0.02 &  0.28 & 8.0 & 3.95 &  & 5150 &  & 4150 \\
0.02 &  0.28 & 9.0 & 4.12 &  & 5050 &  & 3950 \\
0.02 &  0.28 & 10.0 & 4.28 &  & 4850 &  & 3850 \\
0.02 &  0.28 & 11.0 & 4.41 &  & 4850 &  & 3950 \\
\hline
&&&RR Lyrae&&&&\\
\hline
0.001 & 0.245 & 0.58 & 1.87 & 6850 & 6950 & 6350 & 5650 \\
0.001 & 0.245 & 0.64 & 1.67 & 7250 & 6950 & 6750 & 5950 \\
0.02 &  0.28 & 0.51 & 1.69 & & 6850 &  & 5550 \\
0.02 &  0.28 & 0.54 & 1.49 & & 6750 &  & 5850 \\
\hline\hline
\end{tabular}
\end{table*}

\clearpage

\subsection{Bolometric light curves}

\begin{figure}[th]
\centering
\includegraphics[width=\textwidth]{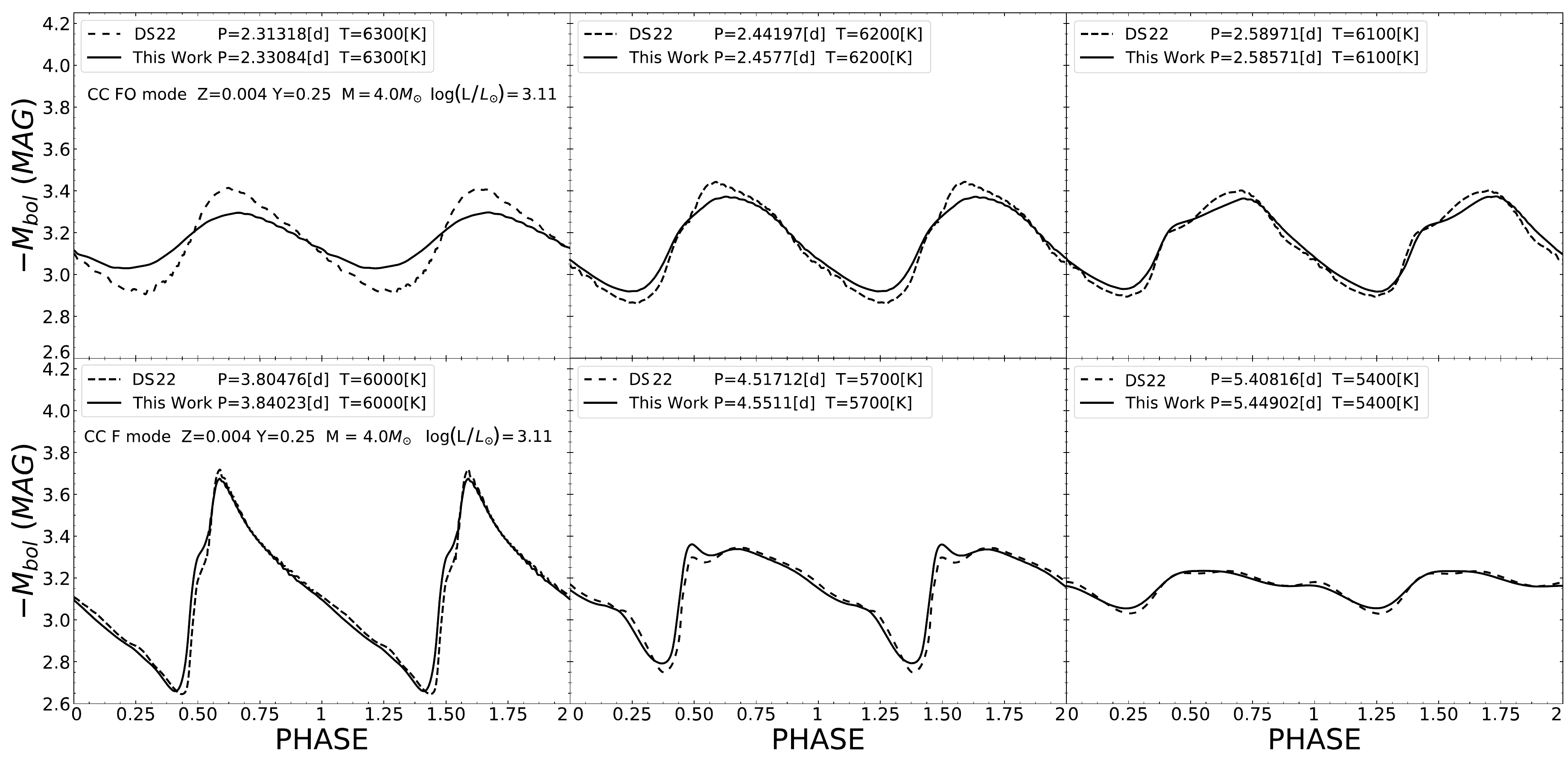}
\caption{Bolometric light curves for FO (upper panels) and F (lower panels) CC pulsators with Z = 0.004 and Y = 0.25, plotted as a function of the pulsation phase. These light curves correspond to selected pulsation models (see labels for the chosen mass and luminosity level) located close to the blue edge (left column), red edge (right column), and central temperature (central column)  of the IS. Dashed lines represent results obtained with the older opacity tables, while solid lines show results with the updated opacity tables. Nonlinear periods (in days) and static effective temperatures (in Kelvin) are labeled for each individual model.}
\label{Fig:lc_cc_smc}
\end{figure}

\begin{figure}[th]
\centering
\includegraphics[width=\textwidth]{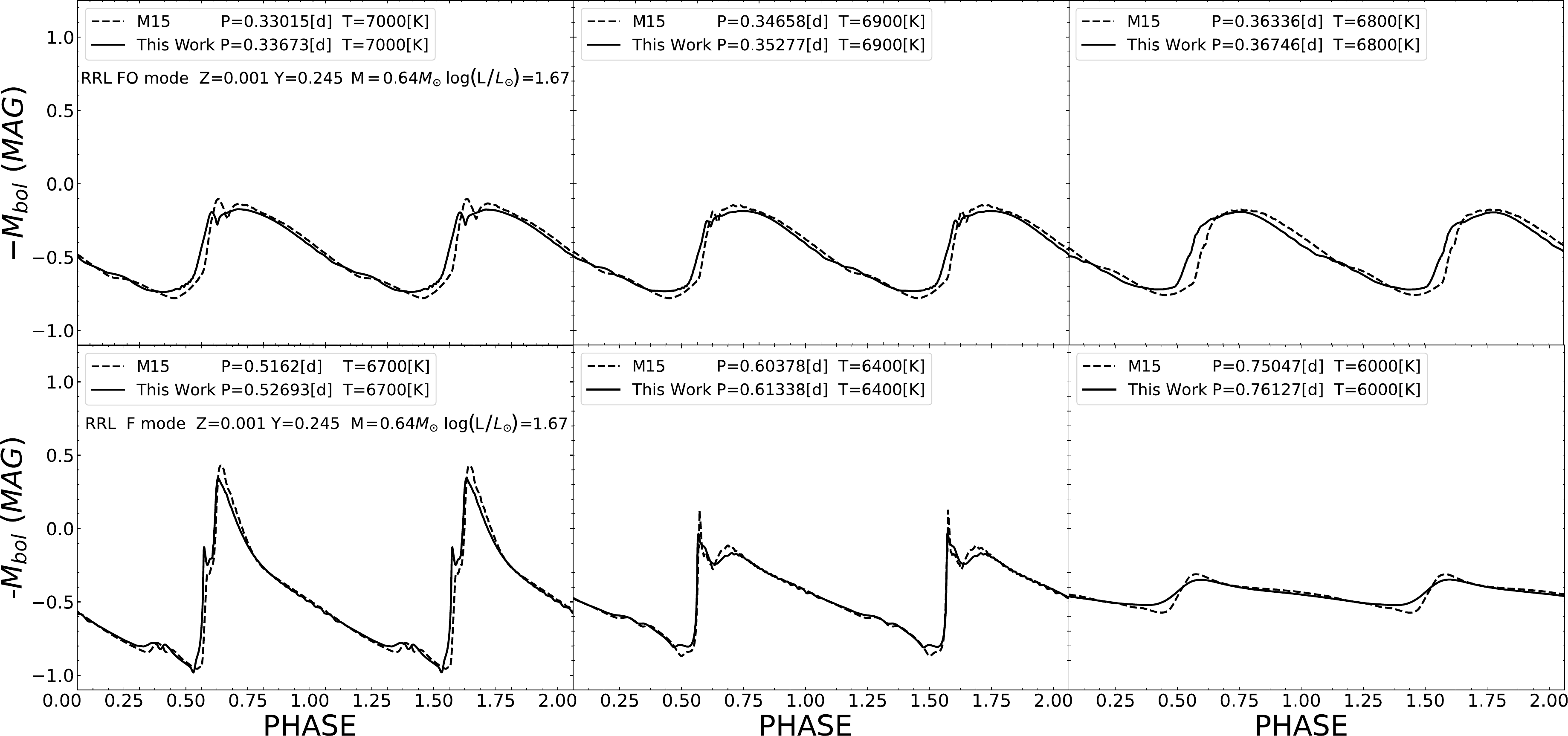}
\caption{As in Fig:\Ref{Fig:lc_cc_smc} but for RRL pulsators with Z=0.001 and Y=0.245.}
\label{Fig:lc_rr_z13}
\end{figure}

We present a theoretical atlas of bolometric light curves for both F and FO modes, derived from the nonlinear computation of full amplitude models. Figure \ref{Fig:lc_cc_smc} shows a sub-sample\footnote{The complete set of bolometric light-curves for CC and RRL models covering the various chemical compositions included in this analysis, is available upon request.} of the predicted light curves (solid lines) compared to previously computed ones (dashed lines) for CC model sequences close to the IS blue and red edge (left and right panels, respectively) and in the middle of the IS (middle panel). 
For the case of RR Lyrae pulsators, a sub-sample of these light curves is shown in  Fig.\ref{Fig:lc_rr_z13}. The light curves, plotted over two consecutive pulsation cycles, are presented as a function of the pulsation phase.

For CC pulsators, there is excellent agreement between previous results and the present ones for fundamental (F) mode pulsators. However, some differences emerge for FO variables. Specifically, the present light curves exhibit a more symmetric behavior and a slightly smaller amplitude - by about $\sim$ 0.1 mag at maximum - compared to previous models. The differences significantly reduce when moving from the the blue edge to the red edge of the IS. A  similar trend is observed for RRL pulsators, although in this case, the differences are on the order of just a few hundredths of a magnitude.

Additionally, we conducted a detailed morphological analysis of these bolometric light curves within the period range of Bump Cepheids. These Cepheids are characterized by the presence of a secondary feature, or "bump," in their light and radial velocity curves, whose position in phase varies with the pulsation period according to a phenomenon known as the Hertzsprung progression (HP) \citep[see e.g.][and references therein]{BMS00}. The position of the bump appears on the descending branch of the curves for periods up to about 9 days, close to the main light/radial velocity maximum for periods between about 9 and 12 days, and at earlier phases for longer periods \citep[see][after this M24, and references therein for a comprehensive discussion on this phenomenon]{Marconi2024}.
The origin of the progression is debated. The most probable explanation involves a resonance between the fundamental mode and the second overtone (\citetalias{Marconi2024}, \citealt{BonoTorn2000, Buchler1986, Buchler1990}).

\begin{figure*}[th]
\centering
\includegraphics[width=\textwidth]{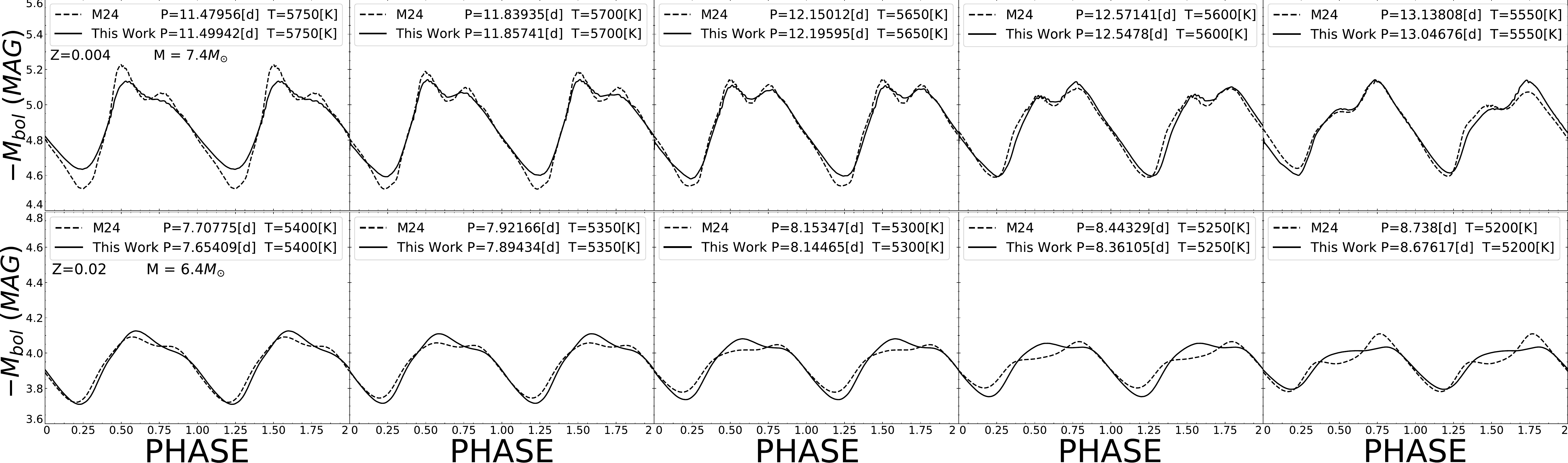}
\caption{Theoretical CC F-mode bolometric light curves, as derived by \citetalias{Marconi2024} (dashed lines) and in the present work (solid lines), along the HP, for Z = 0.004 (upper panels) and Z = 0.02 (lower panels). The adopted stellar mass (in solar units), effective temperature (in kelvin), and inferred pulsation period (in days) are labeled in each panel.}
\label{fig:progression}
\end{figure*}

From our analysis of the evolution of light curve morphology across the HP, for a fixed mass and luminosity, we find that for the lower metallicity adopted in the present analysis (see upper panels of Figure \ref{fig:progression}, the HP center remains consistent with previous investigations based on older opacity tables, as presented in \citetalias{Marconi2024}. Specifically, the period corresponding to the HP center, where the bump is nearly at the same magnitude level as the maximum, remains almost unchanged with the new opacity tables shifting slightly from $P_{HP} = 12.15$ days in \citetalias{Marconi2024} to $P_{HP} = 12.19$ days in the present work. However, at solar metallicity, (see lower panels of Figure \ref{fig:progression}), the HP center shifts towards longer periods compared to previous findings, specifically changing from $P_{HP} = 7.92$ days in the \citetalias{Marconi2024} models to about $P_{HP} = 8.5$ days in the present work. This small variation results from the impact of the opacity update on the morphology of solar metallicity light curves, particularly for models in the middle of the instability strip (see middle panels of Fig.\ref{fig:op_mw_cc}.

Additionally, we note that, with the new opacity tables, the difference in the HP central period between solar and metal-poor models is reduced compared to \citetalias[][]{Marconi2024} findings.

%La differenza nel picco dell'idrogeno riscontrata nelle opacita' per Z=0.02 influenza la forma delle curve di luce inq uesti modelli che si trovano al centro della strip.

\subsection{Pulsation properties: periods, mean magnitudes, and amplitudes}

\begin{figure*}[ht]
\centering
    \vbox{
    \hbox{
    \includegraphics[width=0.45\linewidth]{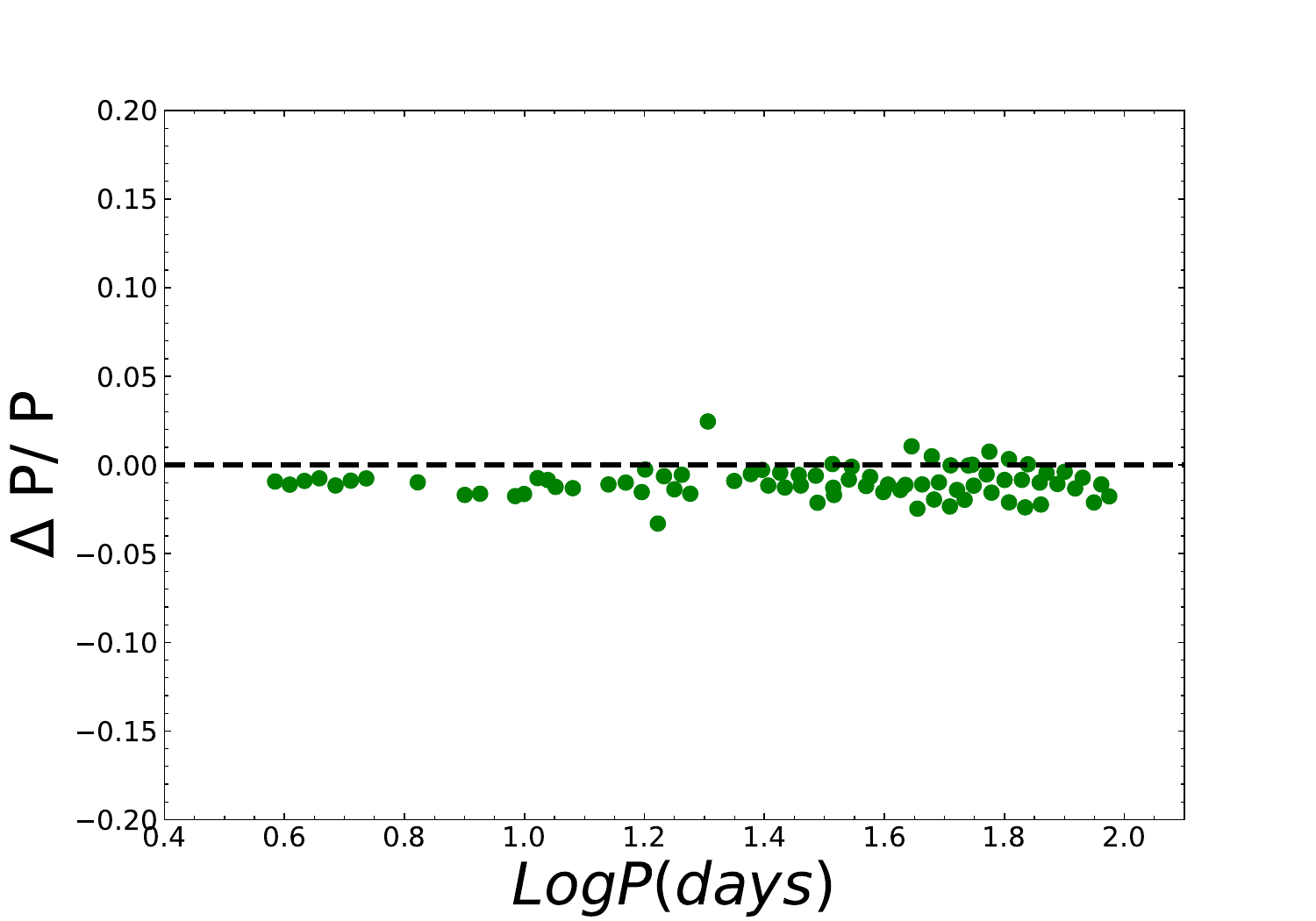}
    \includegraphics[width=0.45\textwidth]{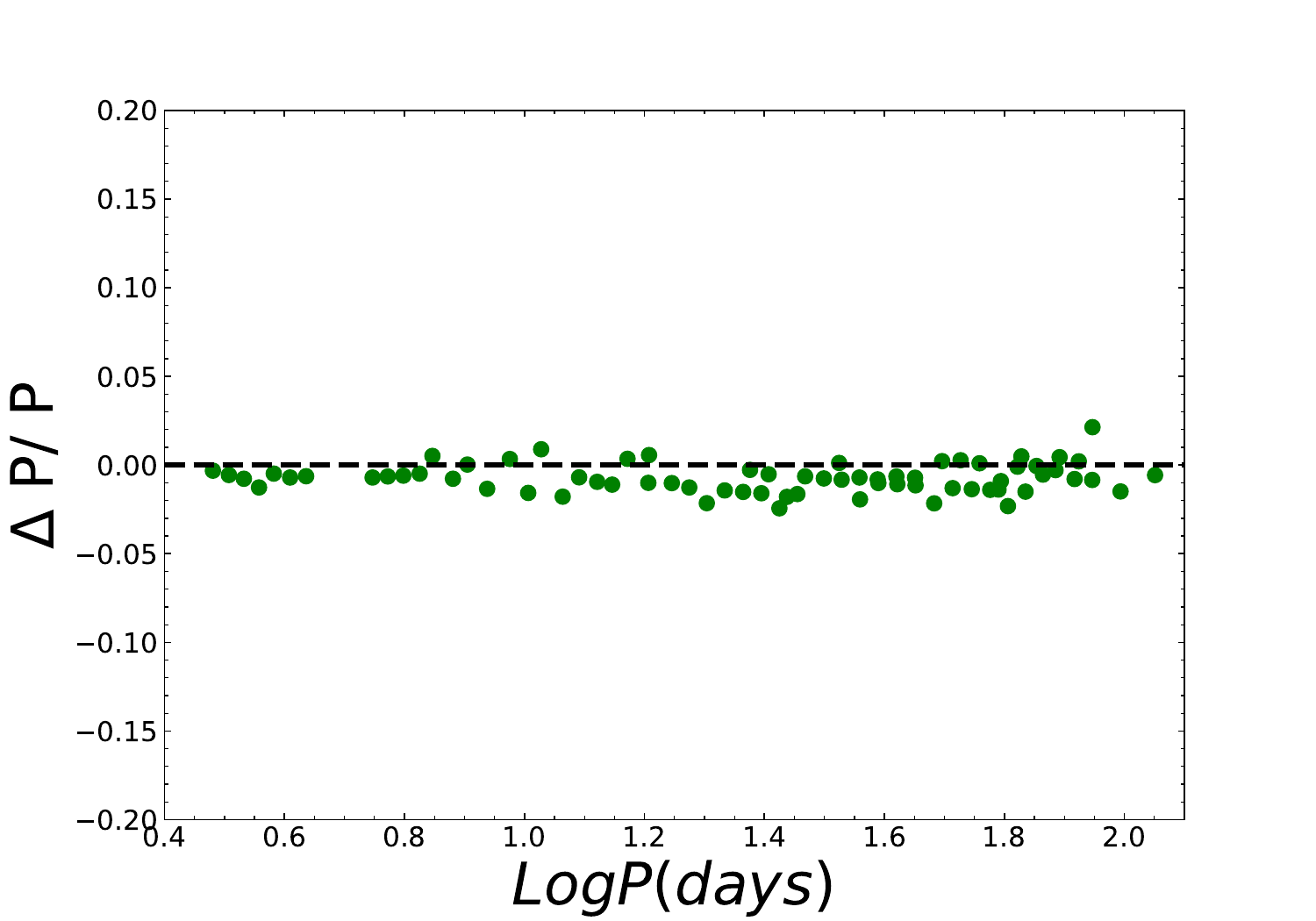}
    }
    \hbox{
    \includegraphics[width=0.45\textwidth]{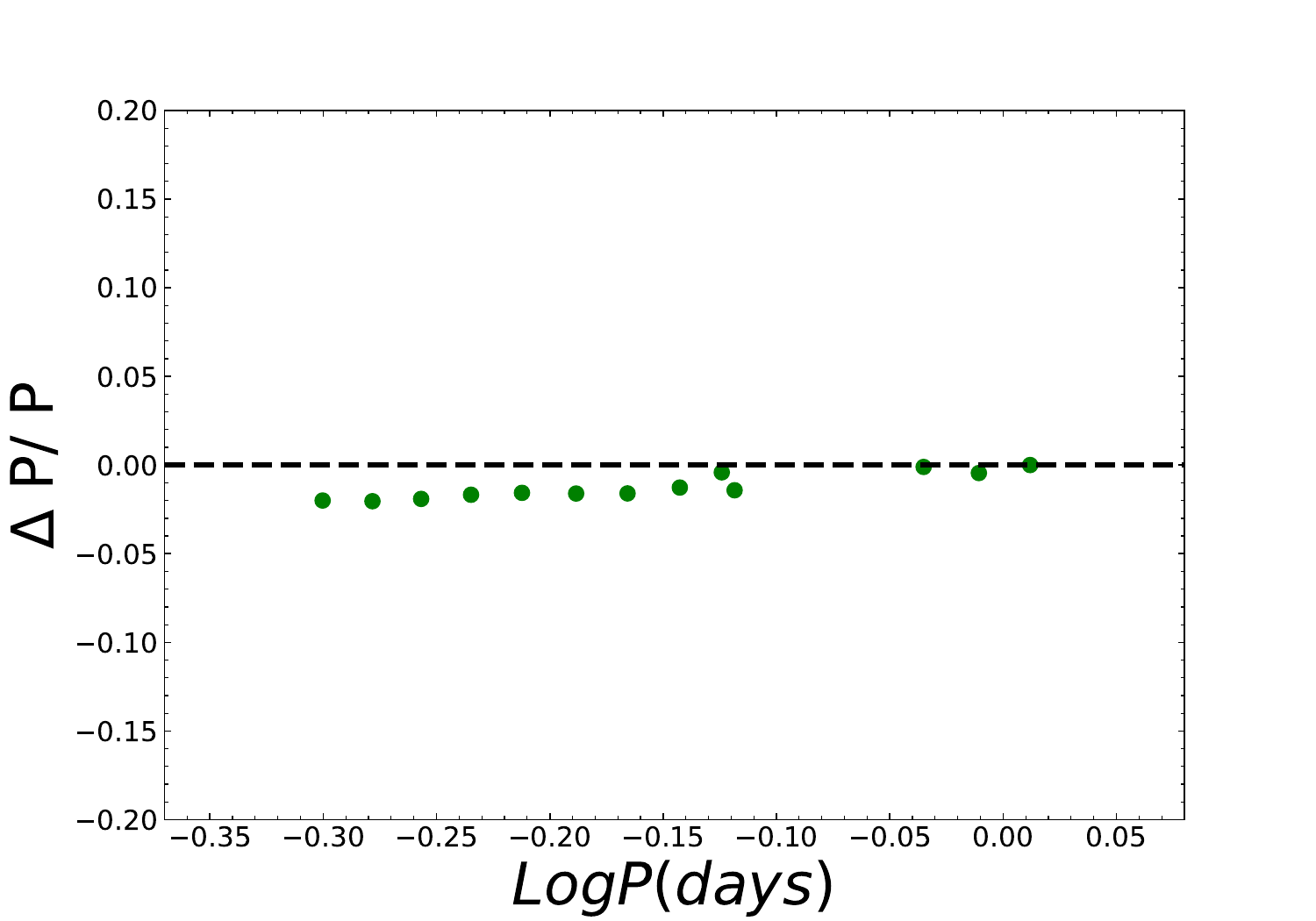}
    \includegraphics[width=0.45\textwidth]{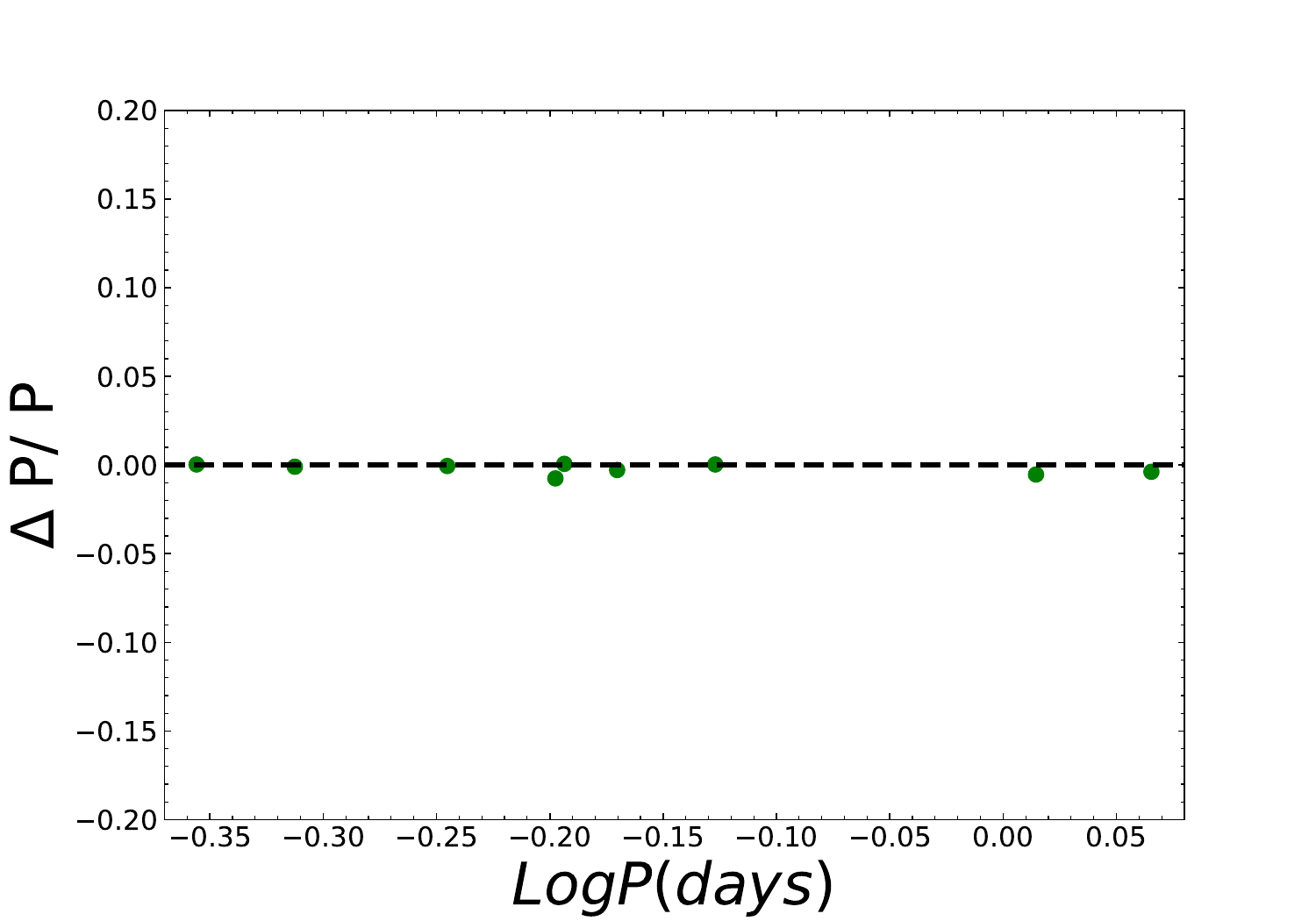}
    }}
    \caption{\label{fig:period_comparison} Upper panels: Relative variations of the pulsation periods obtained from the present models compared to those from \citetalias[][]{Desomma2022} for CCs with $Z=0.004$ and \citetalias{Desomma2020} for CCs with $Z=0.02$. The dashed horizontal line represents the equation y=0. Lower panels: Same as the upper panels but for RR Lyrae pulsators; in this case, the comparison is performed with the estimates provided by \citetalias{Marconi2015}.
    }
\end{figure*}

In the upper panels of Figure \ref{fig:period_comparison}, we show the relative variation of the pulsation periods obtained from the present models compared to those from \citetalias[][]{Desomma2020} and \citetalias[][]{Desomma2022} for CCs at the two metallicities considered in this work, as a function of the logarithm of the pulsation period obtained in the current work when changing the opacity tables. We observe that the variations are all within a few percent and do not show any significant trend with the pulsation period.
A similar result is observed for F-mode RR Lyrae pulsators, as shown in the lower panels of Figure \ref{fig:period_comparison}. This variation estimation was also extended to FO-mode CC and RR Lyrae pulsators with $Z=0.004$ and $Z=0.001$.

\begin{figure*}[ht]
    \hbox{
    \includegraphics[width=0.45\linewidth]{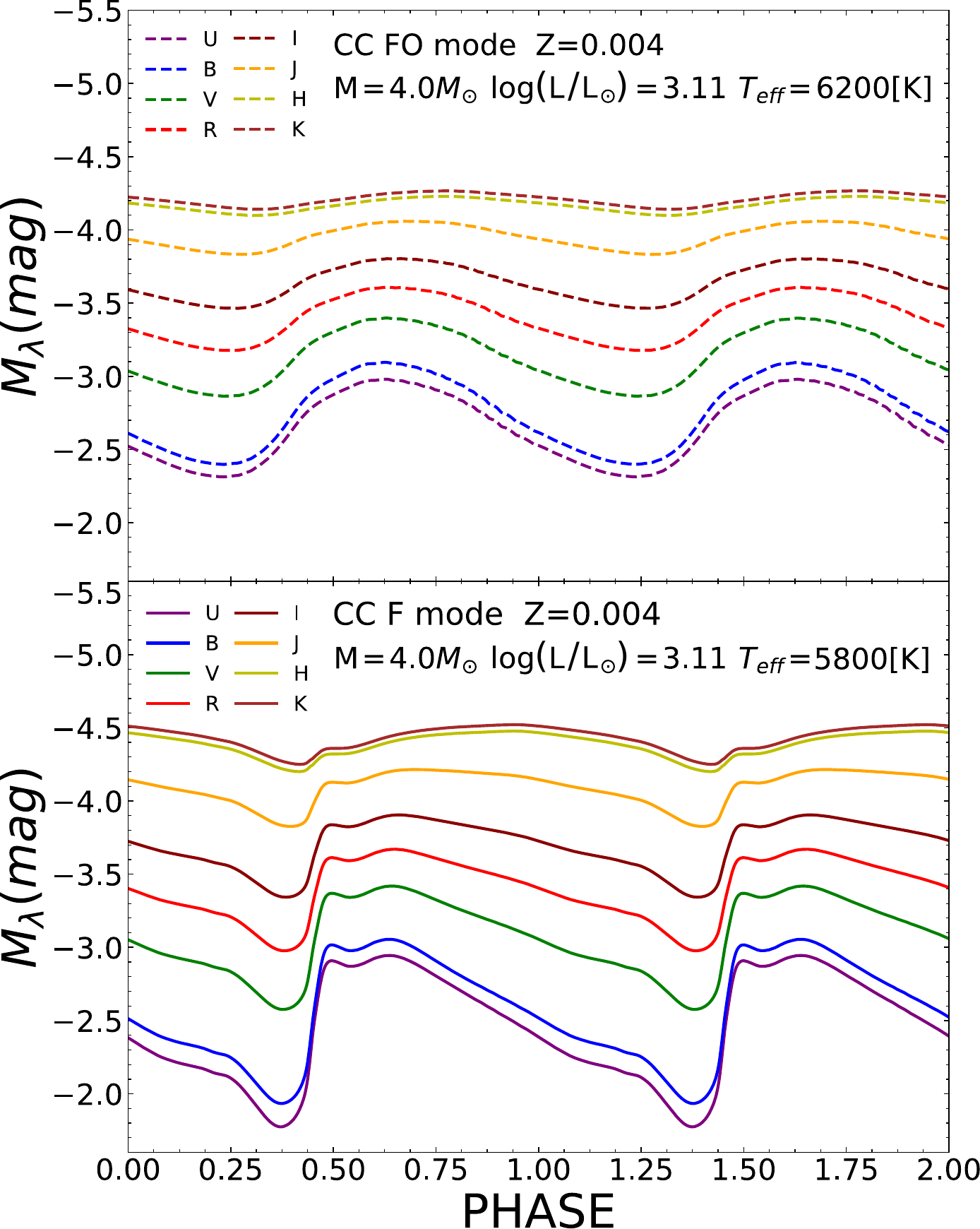}
    \includegraphics[width=0.45\linewidth]{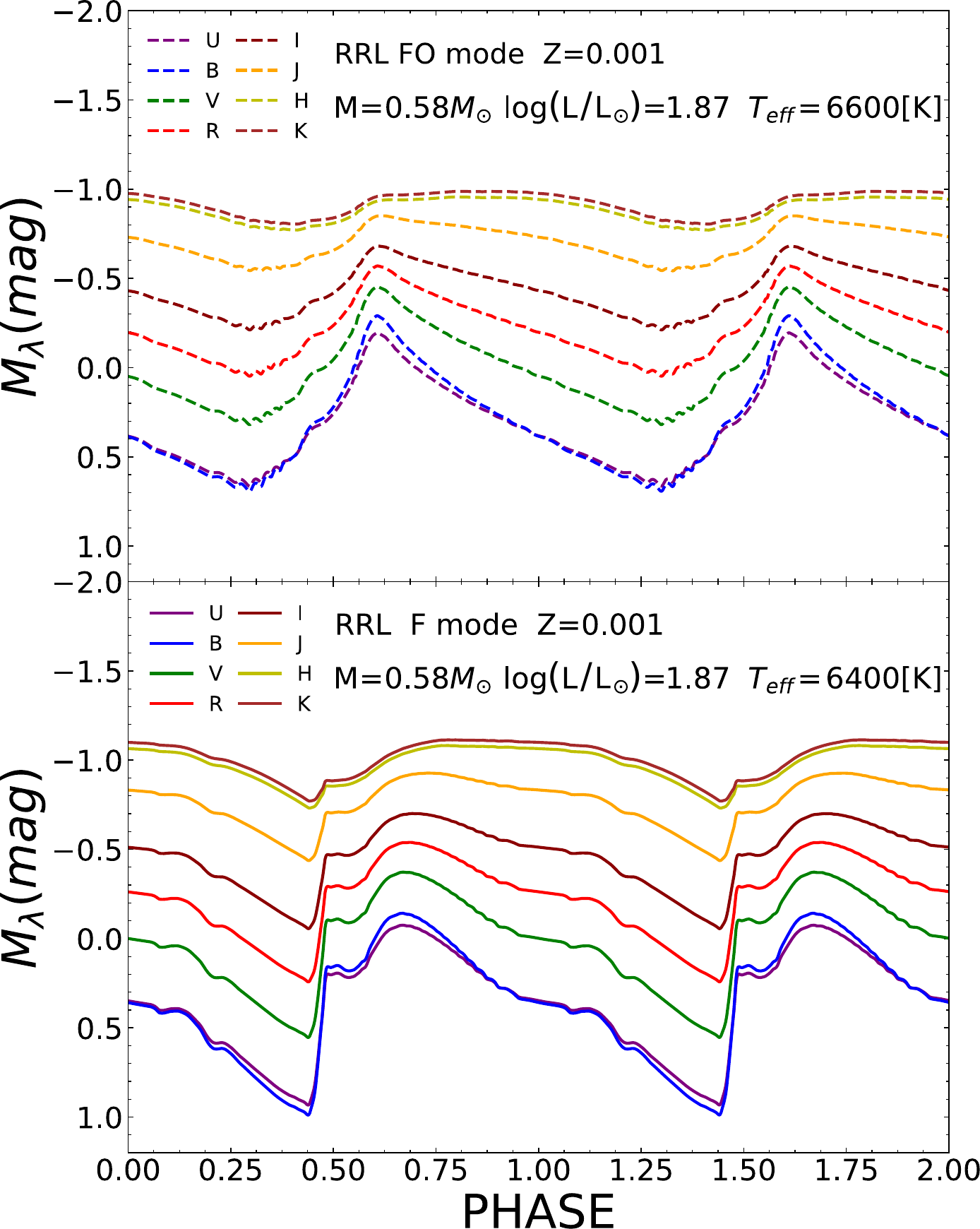}
    }
    \caption{\label{fig:lc_joh_cep_rrl} Left panels: Theoretical light curves in various Johnson-Cousins passbands for a CC F-mode model (bottom panel) and FO-mode model (upper panel), both located at the center of the IS. The metallicity, mass (in solar units), luminosity (in solar units), and effective temperature (in Kelvin) for the selected model are labeled. Right panels: Same as the left panels but for RRL models.}
\end{figure*}

The bolometric light curves\footnote{The complete set of light curves in various photometric systems is available from the authors upon request.} were transformed into the optical UBVRI and NIR JHK Johnson-Cousin photometric bands using the bolometric correction tables by \cite{Chen2019}. A subset of these light curves for both CC and RRL stars is shown in fig.~\ref{fig:lc_joh_cep_rrl}.  This transformation allowed us to derive intensity-weighted mean magnitudes, pulsation amplitudes, and color for all stable pulsation models, in the various photometric bands. We chose to evaluate intensity-weighted mean magnitudes because they are most consistent with static values, which represent the magnitudes and colors the stars would have if they were not pulsating \citep[][]{Bono95, Caputo1999}. \footnote{The intensity-weighted mean magnitudes and amplitudes in the Johnson-Cousins photometric bands for the full grid of F and FO models for CC and RRL stars are available upon request from the authors.}

\subsection{On the predicted CC Period-Wesenheit relation}

The most widely used relation for calibrating the CC-based extragalactic distance scale is the Period-Wesenheit (PW) relation \citep{Madore82}. Unlike the PL relation, which is a statistical relation influenced by the finite width of the instability strip without accounting for position in color within the strip, and unlike the Period-Luminosity-Color (PLC) relation, which includes a color term to reduce the effects of the IS width but is still affected by reddening, the PW relation including a color term, partially correcting for the position of pulsators within the strip, is reddening-free by construction.

The PW relation is expressed as a linear relationship between the logarithm of a star's pulsation period and a Wesenheit function, which combines magnitudes from different photometric bands. The coefficient of the color term in this relation represents the exact ratio between total and selective absorption, ensuring that it is unaffected by reddening.

Using the intensity-weighted mean magnitudes obtained from the bolometric light curves transformed into the various photometric bands, we derived the CC-PW relations for two band combinations: $I, (V-I)$ and $K, (V-K)$. \footnote{The general formulation of the PW relation is: $W(\lambda_{1}, \lambda_{2} - \lambda_{1}) = a + b\times\log(P)$. For the two adopted band combinations, $I, (V-I)$ and $K, (V-K)$, we used the following formulas: $W(I, V-I) = I - 1.38 \cdot (V-I)$ (left panels) and $W(K, V-K) = K - 0.13 \cdot (V-K)$, respectively.} The coefficients of the relations are listed in Table \ref{pw_CC}.

\begin{figure*}[ht]
\centering
    \vbox{
    \hbox{
    \includegraphics[width=0.5\linewidth]{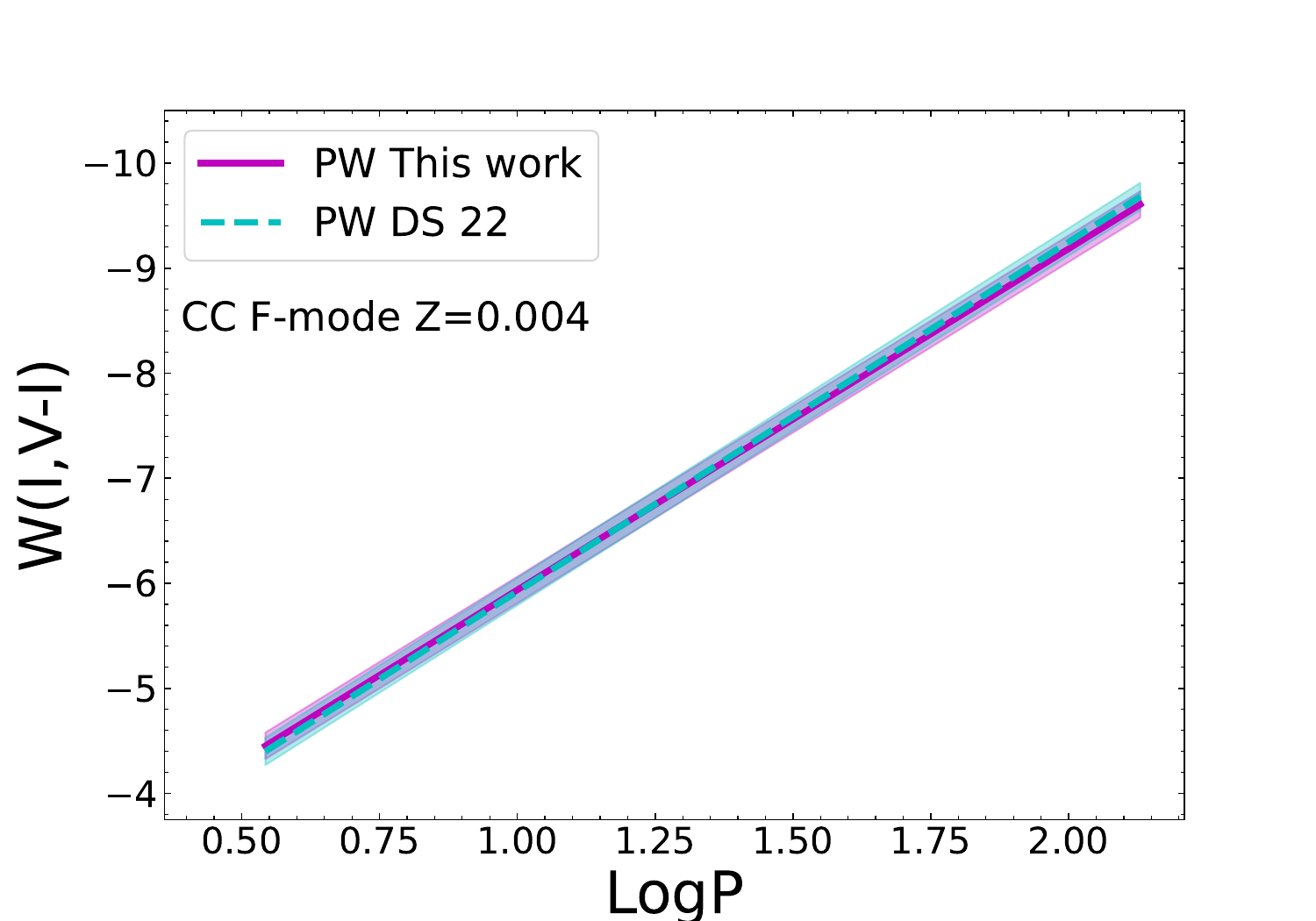}
    \includegraphics[width=0.5\textwidth]{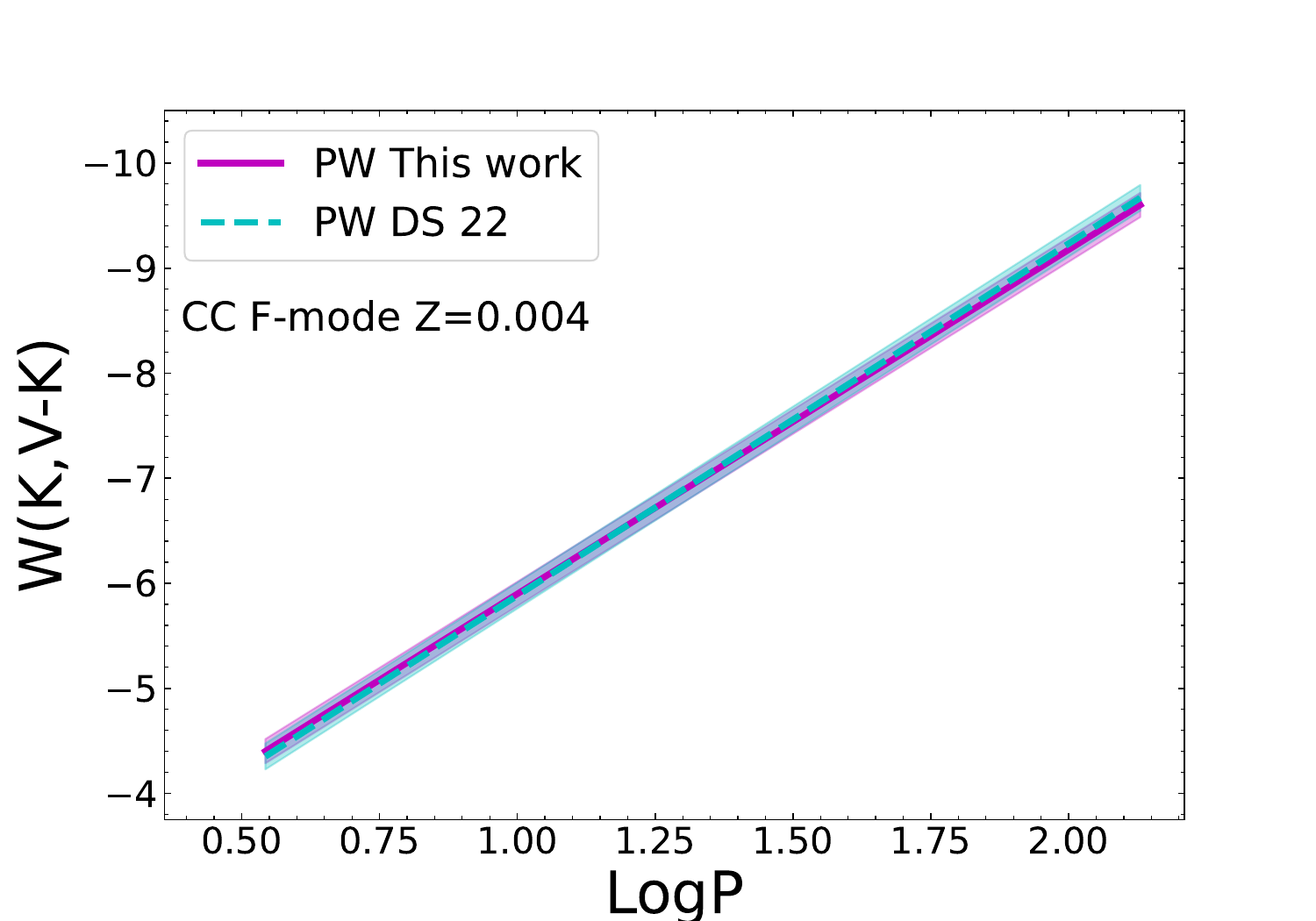}
    }
    \hbox{
    \includegraphics[width=0.5\textwidth]{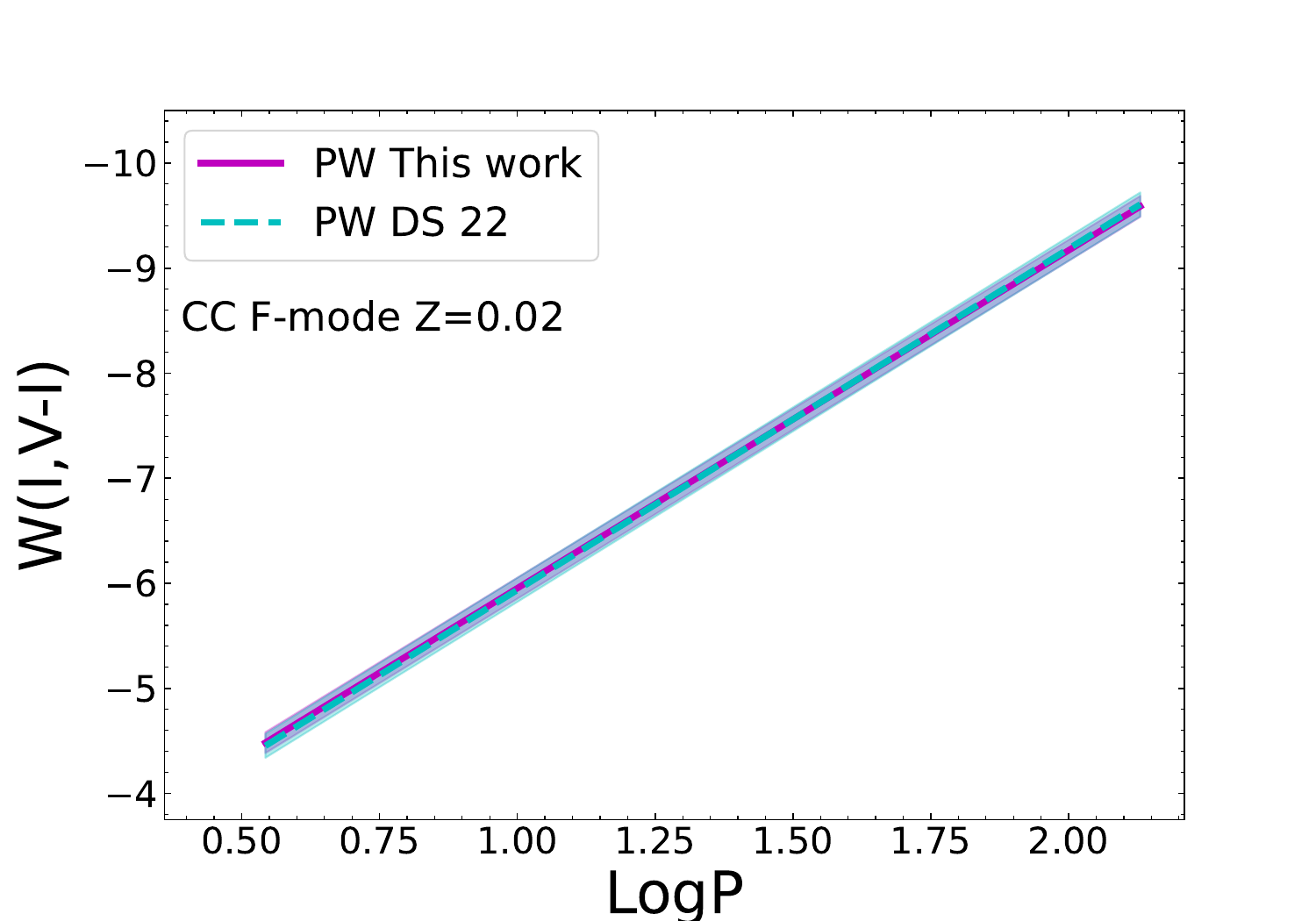}
    \includegraphics[width=0.5\textwidth]{plot_PW_CC_Fmode_mw_I_VI.pdf}
    }}
    \caption{\label{fig:PW} The Wesenheit function, $W$(I, V$-$I) (left panels) and $W$(K, V$-$K) (right panels), plotted as a function of the logarithmic period for F-mode CCs with $Z=0.004$ (upper panels) and $Z=0.02$ (bottom panels).
    }
\end{figure*}

Figure \ref{fig:PW} compares present PW relations with those from \citetalias{Desomma2022} for Z=0.004 and Z=0.02. The colored shaded areas represent the $1\sigma$ errors on these relations. It is evident that the relations are consistent within the errors across the investigated period range and predict a similar metallicity effect, which agrees with most empirical and theoretical evaluations (\citetalias[][]{Desomma2022}, \citealt[][]{Anderson2016, Bhardwaj2023, Breuval2022, Ripepi2021}). This finding indicates that the adopted update in radiative opacity does not significantly impact the theoretical calibration of the distance scale and, consequently, the estimate of the Hubble constant derived from this theoretical approach.

\begin{figure*}[ht]
\centering
    \vbox{
    \hbox{
    \includegraphics[width=0.35\linewidth]{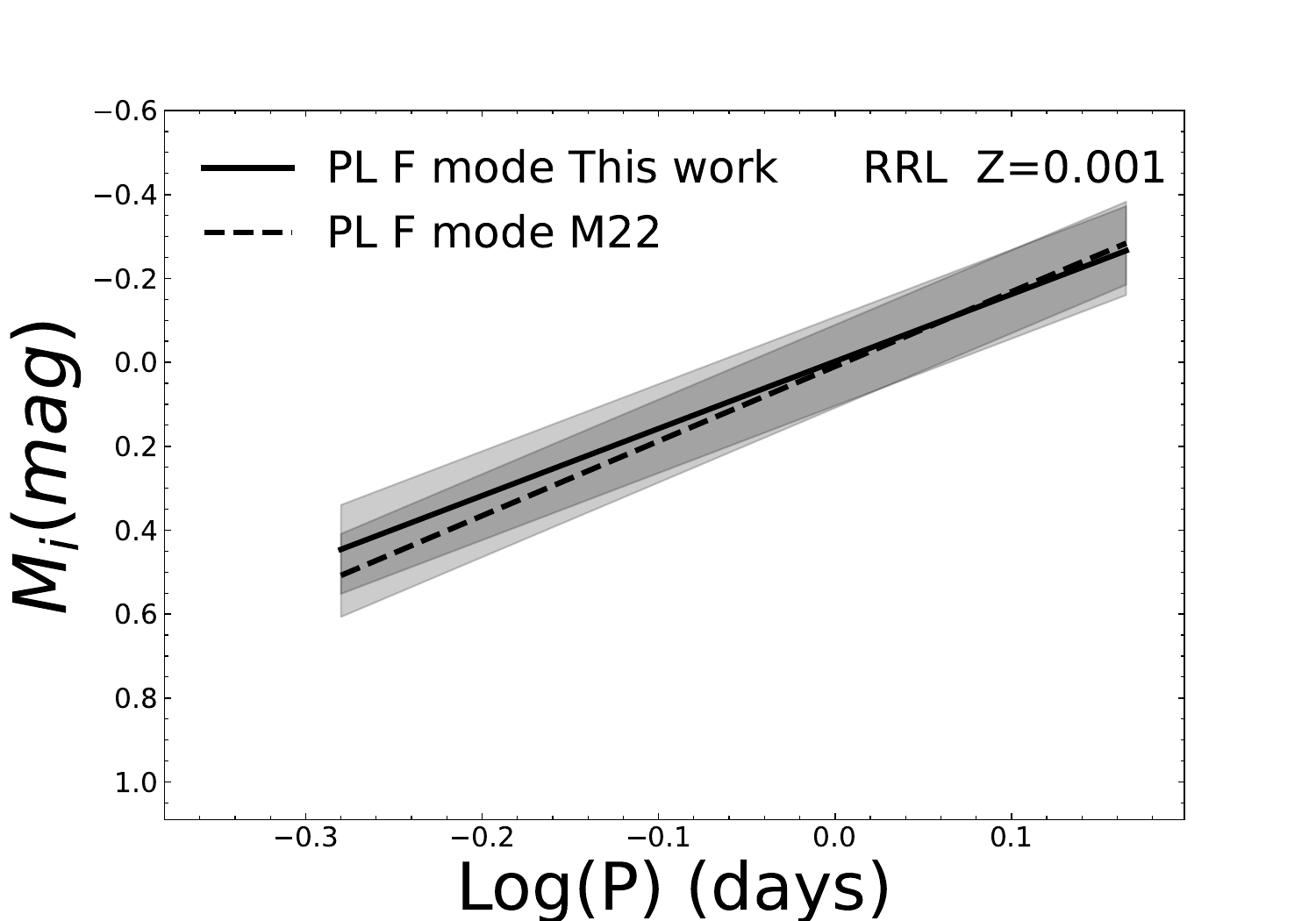}
    \includegraphics[width=0.35\textwidth]{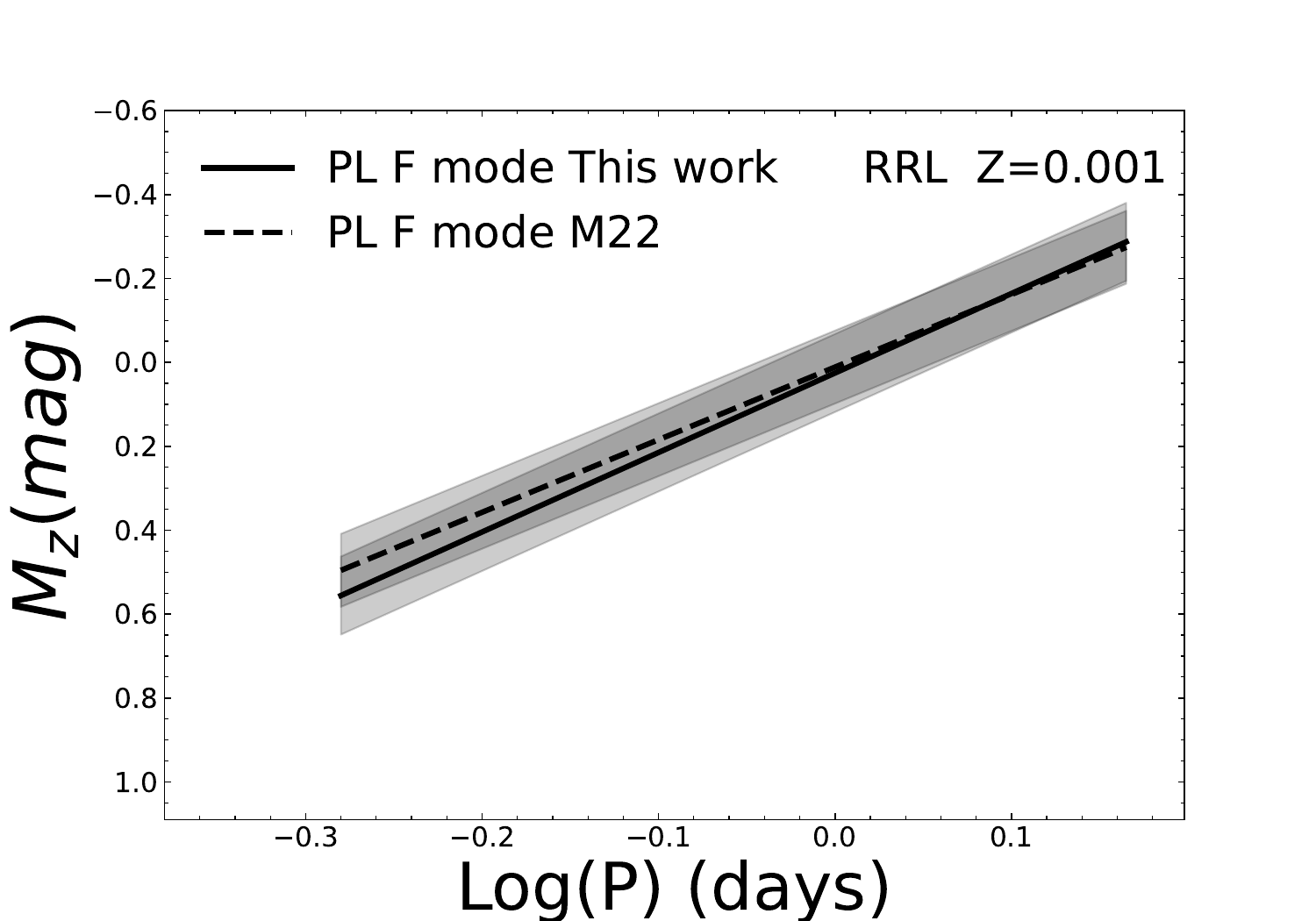}
    \includegraphics[width=0.35\textwidth]{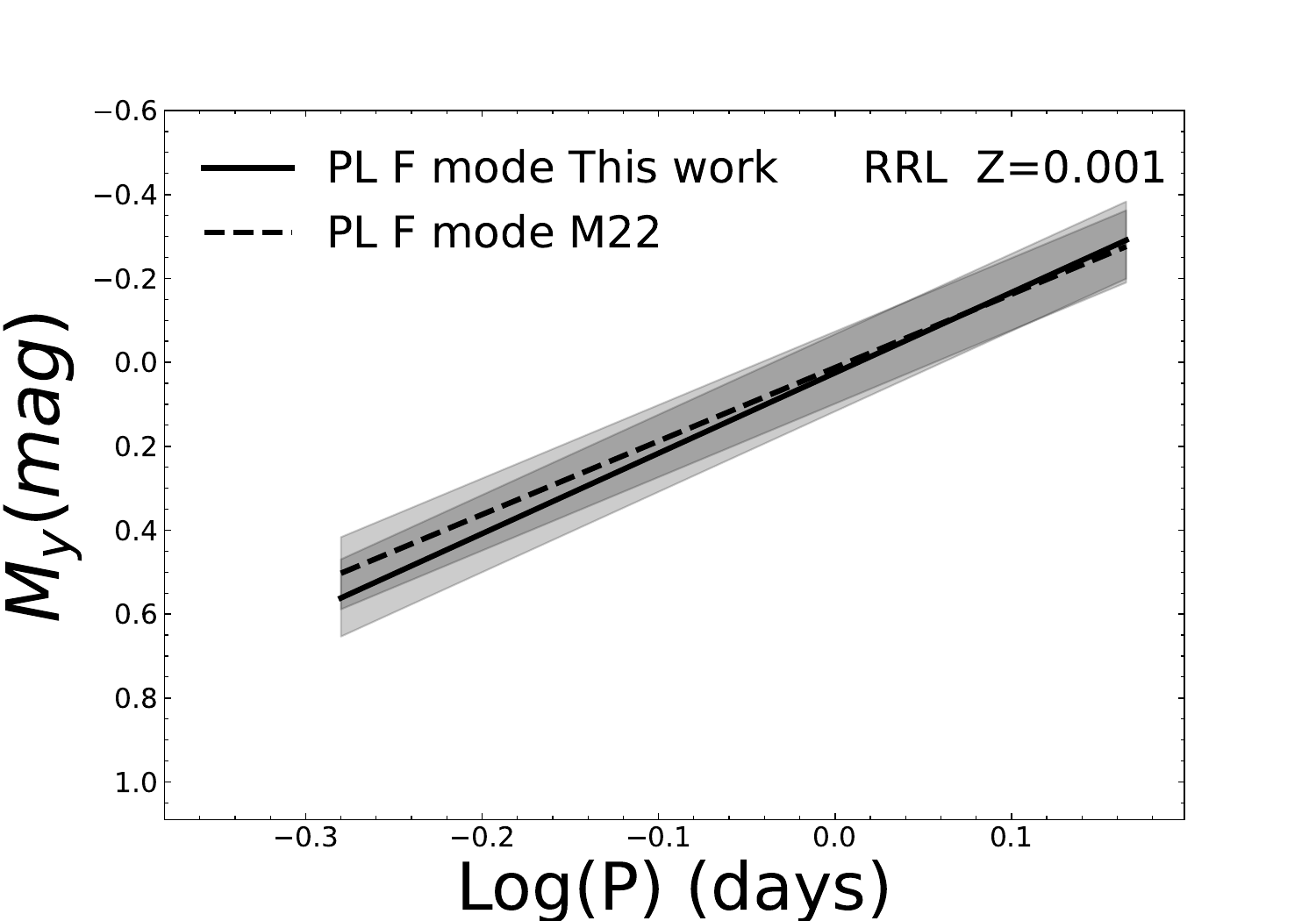}
    }
    \hbox{
    \includegraphics[width=0.35\textwidth]{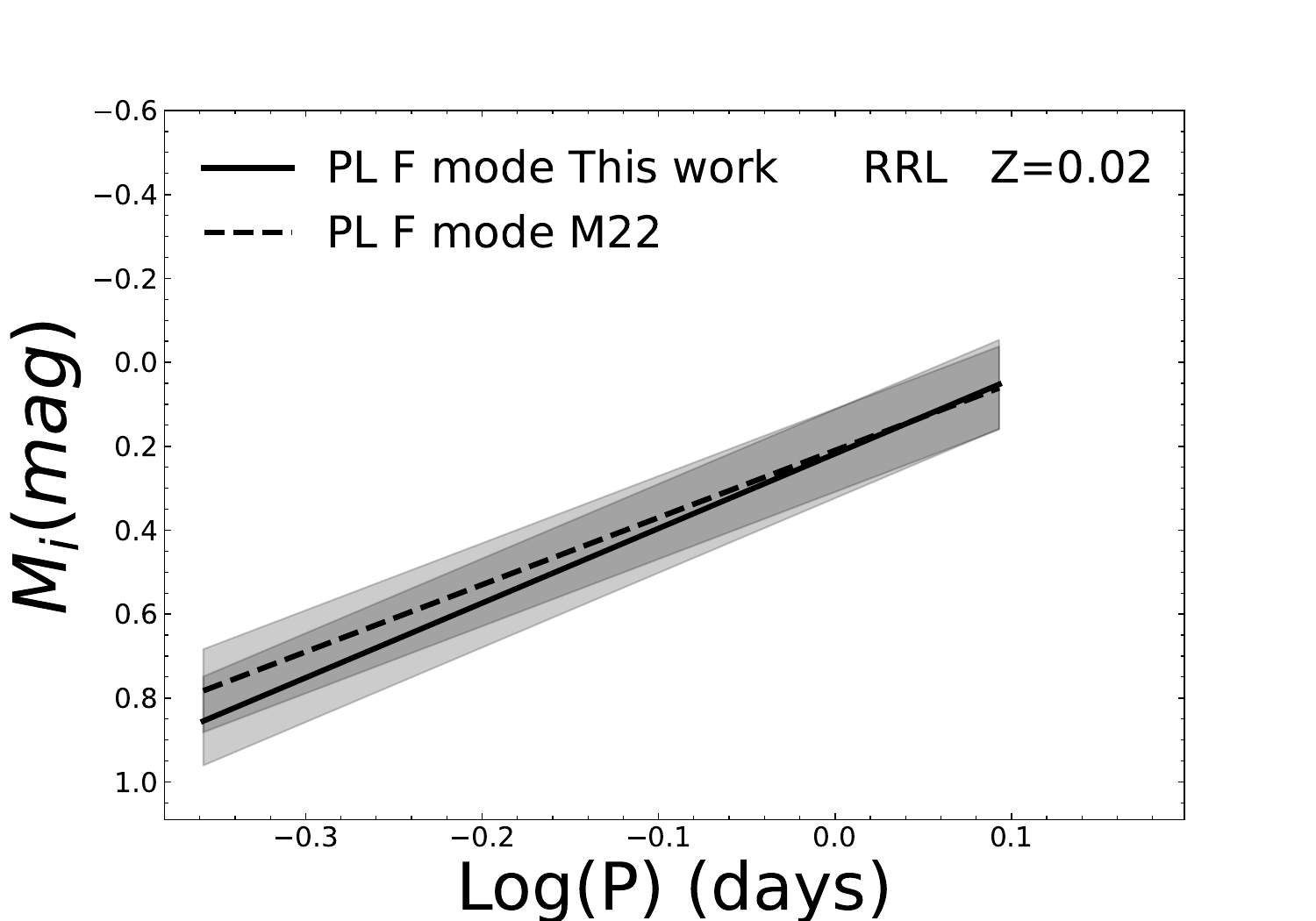}
    \includegraphics[width=0.35\textwidth]{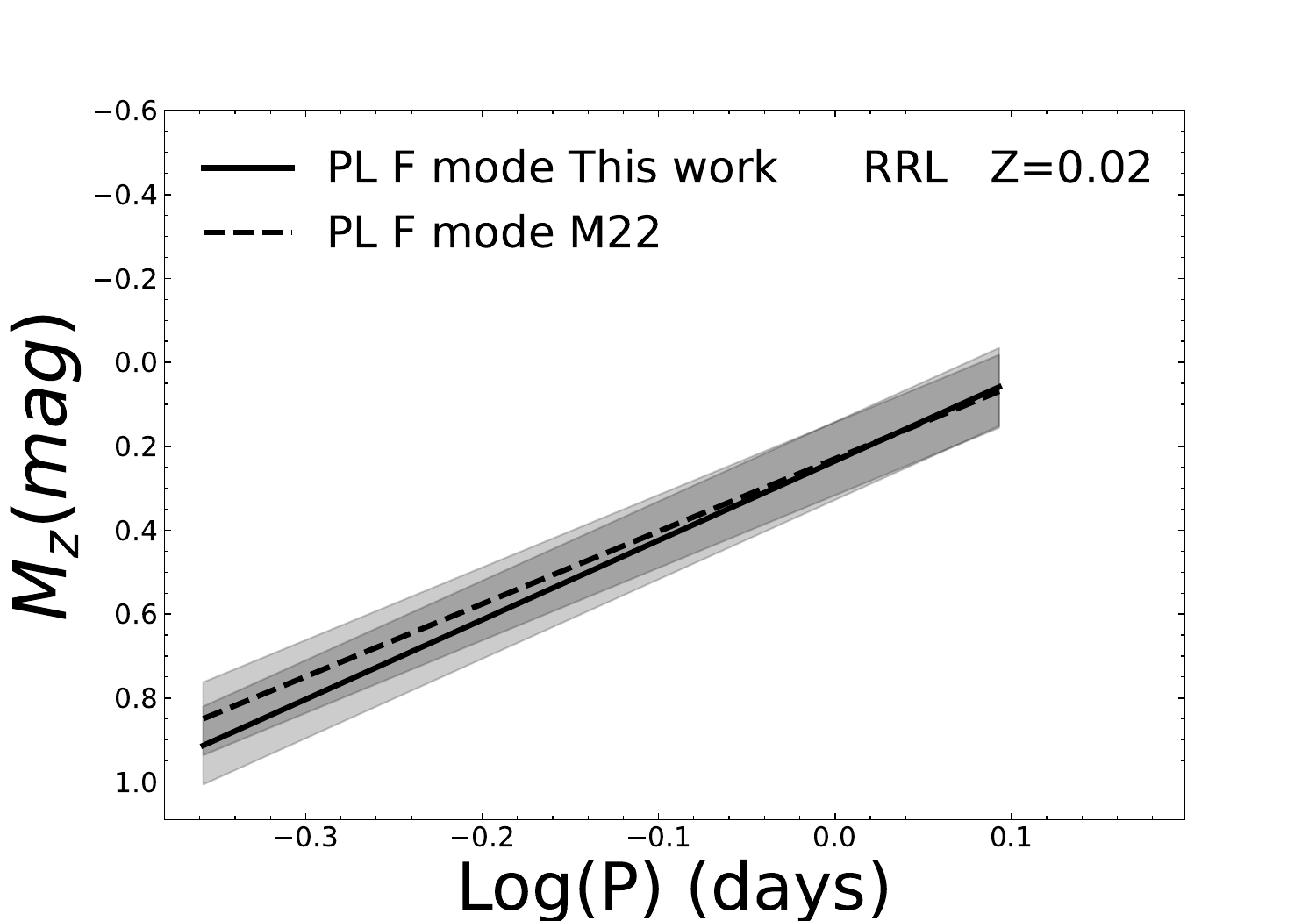}
    \includegraphics[width=0.35\textwidth]{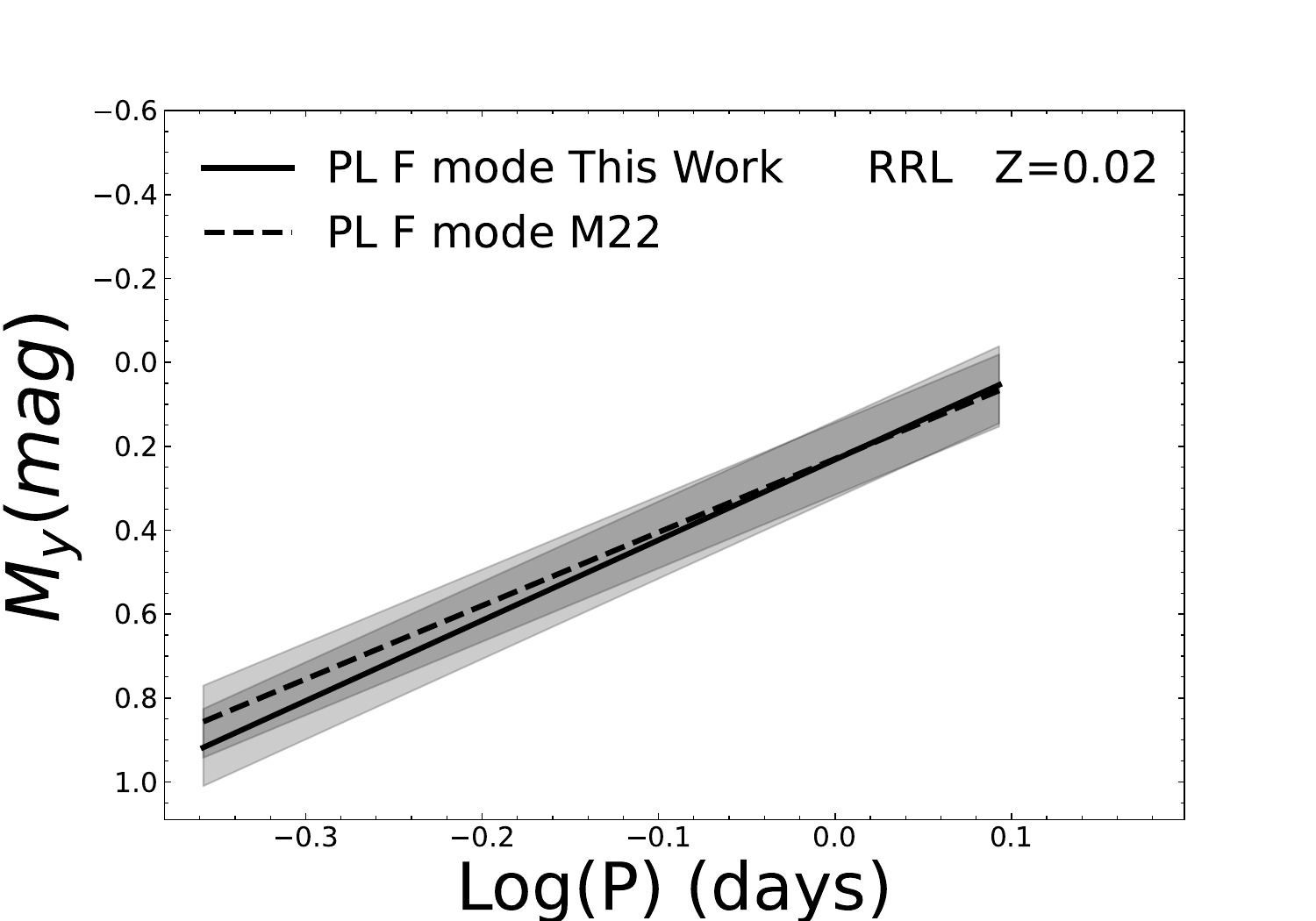}
    }}
    \caption{\label{fig:PL_RRL} Upper panels: Comparison between multifilter PL relations in the Rubin-LSST i, z, and y bands for RR Lyrae stars with $Z=0.001$ derived in the present work and those from \citet[][]{Marconi2022}. Lower panels: Same as the upper panels but for RR Lyrae stars with $Z=0.02$.
    }
\end{figure*}

As well known, RRL stars exhibit tight metal-dependent Period-Luminosity (PL) relations in the near-infrared (NIR) bands \citep[see][and references therein]{Bono2003}. We also converted our bolometric light curves in the Rubinf-LSST photometric system\footnote{The bolometric light curves, as well as the intensity-weighted mean magnitudes and amplitudes in the Rubin-LSST bands for the full grid of F and FO models for CC and RRL stars, are available upon request from the authors.} and derived the PL relations in the
i, z, and y bands of the Rubin LSST for the 2 adopted metal abundances and compared them with those from \citep[][]{Marconi2022}. As shown in Fig. \ref{fig:PL_RRL}, the relations derived in the current paper are consistent within the errors with the \citet[][]{Marconi2022} ones, confirming that, even for RRL stars, the adopted update in radiative opacity does not significantly impact the theoretical calibration of the distance scale.

\begin{table}
\caption{\label{pw_CC} PW coefficients in selected optical and near-infrared Johnson-Cousin filters for F-mode CC models.}
\centering
\begin{tabular}{ccccccccccc}
\hline\hline
Z&Y&Photometric band&a&b&$\sigma_{a}$&$\sigma_{b}$&$\sigma$&$R^2$ \\
\hline
0.004&0.25&I, V$-$I&$-$2.690&$-$3.247&0.048&0.032&0.127&0.991\\
0.004&0.25&K, V$-$K&$-$2.623&$-$3.276&0.044&0.030&0.116&0.992\\
0.02&0.28&I, V$-$I&$-$2.737&$-$3.216&0.037&0.026&0.100&0.995\\
0.02&0.28&K, V$-$K&$-$2.592&$-$3.262&0.033&0.023&0.089&0.996\\
\hline\hline
\end{tabular}
\end{table}

\section{Conclusions and Future Developments}
In this study, we have integrated the latest opacity tables into the Stellingwerf pulsation code and examined their impact on the theoretical models of Classical Cepheids and RR Lyrae stars. Our results indicate that the updated opacities lead to minor differences in predicted pulsation properties such as instability strip topology, light curves, and periods. A detailed morphological analysis of bolometric light curves in the period range of Bump Cepheids confirms the dependence of the phenomenon on metallicity but suggests a reduced effect on the period at the HP center whit the updated opacity tables.

We also derived PW relations for CCs using various band combinations, compared them with previously established relations \citep{Desomma2022}, and found that the new opacity tables do not significantly alter the coefficients. Additionally, we derived near-infrared PL relations for RRL in the Rubin-LSST filters and compared them with previous relations derived by \citet{Marconi2022}, finding consistent results within the errors. These findings suggest that no significant effect is expected on our theoretical calibration of the distance scale based on these pulsating stars.

Although the changes are minor, using the most accurate physical inputs for stellar modeling is an important step toward the development of a unified tool for stellar evolution and pulsation.
In future work, we plan to extend this analysis to a broader range of stellar models and incorporate additional physical processes as they become available in stellar evolution and pulsation modeling. We also aim to integrate pulsation computations into updated evolutionary codes adopting the same physical ingredients to enable real-time modeling of both stellar evolution and pulsation properties. This approach promises a more self-consistent framework for understanding variable stars and paves the way for more accurate insights into the physics of pulsating stars and their role as standard candles.

\section*{Acknowledgements}
We thank the Referee for their insightful and constructive feedback on this paper.
This project has received funding through the INAF-ASTROFIT fellowship and the PRIN MUR 2022 project (code 2022ARWP9C) 'Early Formation and Evolution of Bulge and Halo (EFEBHO),' PI: Marconi, M., funded by the European Union – Next Generation EU, Large Grant INAF 2023 MOVIE.

G.D.S. and M.M. thank the ISSI International Team project SHoT: The Stellar Path to the $H_{0}$ Tension in the Gaia, TESS, LSST, and JWST Era (PI: G. Clementini). G.D.S. also thanks the Istituto Nazionale di Fisica Nucleare (INFN), Naples section, for specific initiatives QGSKY and Moonlight2, as well as Gaia DPAC funds from the Istituto Nazionale di Astrofisica (INAF) and the Agenzia Spaziale Italiana (ASI) (PI: M.G. Lattanzi).

This paper is based on work supported by COST Action CA21136, Addressing Observational Tensions in Cosmology with Systematics and Fundamental Physics (CosmoVerse), funded by COST (European Cooperation in Science and Technology).

\bibliography{desomma_main_apj}{}
%%\bibliography{sample63}{}
\bibliographystyle{aasjournal}

\end{document}